\documentclass[longauth,colorlinks=true,linkcolor=black,citecolor=blue,urlcolor=blue]{aa}  
\usepackage{orcidlink}
\usepackage{multirow}
\usepackage{graphicx}
\usepackage{ulem}
\usepackage{xcolor}

\usepackage[version=4]{mhchem}
\usepackage{txfonts}
\begin{document} 

    \title{MINDS. Strong oxygen depletion in the inner regions of a very low-mass star disk?}

   \author{Jayatee Kanwar \orcidlink{0000-0003-0386-2178}
          \inst{1,2,3} \thanks{Current affiliation: Department of Astronomy, University of Michigan, 1085 S. University Ave, Ann Arbor, MI 48109, USA}
          \and
          Inga Kamp \orcidlink{0000-0001-7455-5349} \inst{1} \and
          Peter Woitke \orcidlink{0000-0002-8900-3667} \inst{2} \and
          Ewine F. van Dishoeck \orcidlink{0000-0001-7591-1907} \inst{4,5} \and
          Thomas Henning \orcidlink{0000-0002-1493-300X} \inst{6} \and
          Yao Liu \orcidlink{0000-0002-7616-666X} \inst{7} \and
          Till Kaeufer \orcidlink{0000-0001-8240-978X} \inst{1,2,3,8,14}\and
          Beno\^{i}t Tabone \orcidlink{0000-0002-1103-3225} \inst{11} \and
          Manuel G\"udel \orcidlink{0000-0001-9818-0588} \inst{9,10} \and
          David Barrado \orcidlink{0000-0002-5971-9242} \inst{12} \and
          Aditya M. Arabhavi \orcidlink{0000-0001-8407-4020} \inst{1} \and
          Riccardo Franceschi \orcidlink{0000-0002-8889-2992} \inst{13} \and
          Marissa Vlasblom \orcidlink{0000-0002-3135-2477} \inst{4}
          }

   \institute{Kapteyn Astronomical Institute, University of Groningen, P.O. Box 800, 9700 AV Groningen, The Netherlands\\
              \email{jkanwar@umich.edu} \and
    Space Research Institute, Austrian Academy of Sciences, Schmiedlstr. 6, A-8042, Graz, Austria \and 
    Institute for Theoretical Physics and Computational Physics, Graz University of Technology, Petersgasse 16, 8010 Graz, Austria \and
    Leiden Observatory, Leiden University, PO Box 9513, 2300 RA Leiden, the Netherlands \and
    Max-Planck Institut f\"{u}r Extraterrestrische Physik (MPE), Giessenbachstr. 1, 85748, Garching, Germany \and
    Max-Planck-Institut f\"{u}r Astronomie (MPIA), K\"{o}nigstuhl 17, 69117 Heidelberg, Germany \and
    School of Physical Science and Technology, Southwest Jiaotong University, Chengdu 610031, China \and
    SRON Netherlands Institute for Space Research, Niels Bohrweg 4, NL-2333 CA Leiden, the Netherlands \and
    Dept. of Astrophysics, University of Vienna, T\"urkenschanzstr. 17, A-1180 Vienna, Austria \and
    ETH Z\"urich, Institute for Particle Physics and Astrophysics, Wolfgang-Pauli-Str. 27, 8093 Z\"urich, Switzerland \and
    LESIA, Observatoire de Paris, Université PSL, CNRS, Sorbonne Université, Univ. Paris Diderot, Sorbonne Paris Cité, 5 place Jules Janssen, 92195 Meudon, France\and
    Centro de Astrobiolog\'ia (CAB), CSIC-INTA, ESAC Campus, Camino Bajo del Castillo s/n, 28692 Villanueva de la Ca\~nada, Madrid, Spain \and
    Universit\'e Paris-Saclay, CNRS, Institut d’Astrophysique Spatiale, 91405, Orsay, France \and
    Department of Physics and Astronomy, University of Exeter, Exeter EX4 4QL, UK}

   \date{}

\abstract{JWST is discovering a plethora of species in planet-forming disks around very low-mass stars such as \ce{C2H2}, \ce{C6H6}, \ce{C4H2}, \ce{CH3} etc. The column densities of these species retrieved from 0D slab models are very large, e.g. of the order of $10^{20}$\,cm$^{-2}$. This is indicating a carbon-dominated chemistry in a gas with a high C/O ratio. The disk around 2MASS-J1605321-1993159 (M4.5) is one such source showing a molecular pseudo-continuum of \ce{C2H2}. Still two oxygen-bearing molecules, CO and \ce{CO2} are also detected in this source.}
{We aim to take the next step beyond 0D slab models to interpret the spectrum. We examine whether 2D thermo-chemical disk models can produce the large inferred column densities of \ce{C2H2} in the inner regions of the disk and produce a pseudo-continuum in the mid-IR spectrum. We also want to constrain  whether depletion of oxygen or enrichment of carbon is causing the high C/O ratio triggering a carbon-dominated chemistry.}
{We utilize the radiative thermo-chemical disk model P{\tiny RO}D{\tiny I}M{\tiny O} to identify a disk structure which is capable of producing the observed molecular emission of species such as CO, \ce{CO2}, \ce{C2H2}, and \ce{H2O} simultaneously. The spectrum is generated using the fast line tracer FLiTs. We derive the gas temperature $\langle T \rangle$, column density $\langle$ log$_{\rm {10}} N\rangle$ and the emitting area $\langle r_{\rm{1}} - r_{\rm{2}} \rangle$ for these molecules from the 2D disk model and compare them to the parameters retrieved originally from 0D slab models. We use the different effect that changing the O or C abundance has on \ce{CO} and \ce{C2H2} respectively to discriminate between O depletion and C enhancement.}
{We find that a disk structure characterised by the presence of a gap can best explain the observations. The inner disk is strongly depleted in dust, especially small grains ($<5\,\mu$m), and elemental oxygen, leading to a large C/O ratio. This is required to produce a molecular pseudo-continuum of \ce{C2H2} and at the same time a relatively weak \ce{CO} emission. The P- and R-branch of \ce{C2H2} probe deeper layers of the disk whereas the Q-branch probes mostly the surface layers. The combined emission of CO and \ce{CO2} puts strong constraints on the gap's location (0.1-0.5\,au) given a disk gas mass. We report here also the new detection of the CO $\nu$= 2$\rightarrow$1 transition in the JWST spectrum.}
{2D thermo-chemical disk models are able to produce the observed molecular pseudo-continuum of \ce{C2H2}. We find that the combination of different species emission in the JWST spectra can be used to discriminate between different scenarios such as O-depletion, C-enhancement or both and offers the potential to extract spatial substructure at scales smaller than $\sim\,1$\,au.}
   
   \keywords{(stars:) brown dwarf -- protoplanetary disks -- methods: numerical -- infrared: planetary systems -- line: formation -- astrochemistry
               }
\authorrunning{J.Kanwar et al.}
   \maketitle
%
\section{Introduction}
Our galaxy harbors a significant number ($\sim$20\%) of very low-mass stars (VLMSs) \citep[<\,0.3\,M$_{\odot}$]{Liebert1987} \citep{Kirk2012}. They are known to have high occurrence rates of terrestrial planets \citep{Sabotta2021, Schlecker2022}. Disks around VLMS are faint and, hence, difficult to observe. \citet{Pascucci2009, Pascucci2013} observed the gas in the inner regions of these faint sources with the \textit{Spitzer} Space Telescope and detected molecules such as \ce{C2H2} and \ce{HCN}. These inner regions are warm (200-1000\,K) and dense (10$^{8}$-10$^{15}$\,cm$^{-3}$) and are the sites for planet formation \citep{Henning2013, Walsh2015,Ormel2017}. With the advent of the James Webb Space Telescope (JWST), we gain unprecedented insight into these regions due to its higher spectral resolution and sensitivity, albeit without spatial resolution. 2MASS-J1605321-1993159 \citep[hereafter J160532,][]{Tabone2023}, ISO-Chal 147 \citep{Arabhavi2023}, Sz28 \citep{Kanwar2024}, Cha H$\alpha$ 1 (Calderon et al., in prep), and J043814 \citep{Perotti2025} are disks around VLMSs that have now been observed with JWST/MIRI-MRS \citep{miri_rieke2015PASP, Wells2015, Wright2015,Wright2023,  Rigby2023, Labiano2021} as a part of MIRI mid-Infrared Disk Survey (MINDS) program \citep{Henning2024, Kamp2023}. \cite{Arabhavi2025b} report detection of various hydrocarbons in many VLMSs disks in the MINDS sample. Sz114 \citep{Xie2023} is another disk around an M5 star observed with JWST. While Sz114 and J043814 are both detected with the Atacama Large Millimeter/Submillimeter Array (ALMA), J160532, ISO-Cha1 147 and Sz28 are not detected with ALMA in the continuum \citep{Barenfeld2016}, indicating that these disks have a very low dust mass. Disks around J160532, ISO-Cha1 147 and Sz28 show emission from a variety of hydrocarbons at mid-infrared wavelengths, indicating a high C/O ratio in their inner regions. However, whether this is due to the depletion of oxygen or the enrichment of carbon is yet to be determined. Investigating the gas composition within these inner disks can help to understand the initial ingredients for planet formation.

The disk around the M4.5 star, J160532 \citep[$M_{\rm{\star}}$=0.16\,$M_\odot$,][]{Pascucci2013} is one such source with a high C/O elemental ratio in the gas. It is a member of $\beta$\,Sco \citep[7.6\,Myr,][]{Ratzenbock2023} in the Upper Scorpius star-forming region which is 3-20\,Myrs old \citep{Ratzenbock2023} at a distance of 152\,pc \citep{Gaia}. J160532 is accreting with rate between 10$^{-10}$ and 10$^{-9}$ M$_\sun$\,yr$^{-1}$ consistent with the strength of molecular hydrogen lines in the spectrum \citep{Rigliaco2015,Riccardo2024}. Its estimated UV luminosity [log($L_{\rm{FUV}}$)] is -3.59\,L$_{\odot}$ and has only an upper limit on X-rays [log($L_{\rm{X}}$)] of 28.8 ergs s$^{-1}$. The disk shows no 10\,$\mu$m silicate feature in \textit{Spitzer} nor JWST-MIRI observations, indicating that either the grains are large or that dust is largely absent from the inner warm disk. The peculiar shape of the spectral energy distribution (SED) between 5 and 20\,$\mu$m has been explained by \cite{Tabone2023} with a molecular pseudo-continuum of \ce{C2H2}. They used LTE 0D slab models to identify and analyse the emission in the JWST-MIRI/MRS spectrum of J160532 and found the pseudo-continuum to be caused by optically thick emission of \ce{C2H2} with column densities of $\sim$10$^{20}$\,cm$^{-2}$ and temperatures of $\sim$500\,K. In addition, they required an optically thin colder component of \ce{C2H2} (400\,K) producing the prominent Q-branch. \citet{Tabone2023} attributed such high column densities of \ce{C2H2} to the lack of grains in the inner disk, thus allowing us to look very deep into layers close to the midplane. \cite{Tabone2023} explained these detections by invoking a gap that stops the flow of icy grains into the inner disk resulting in depletion of oxygen in the inner disk. This spectrum also shows the first detections of \ce{C6H6} and \ce{C4H2} in the inner regions of a planet-forming disk and the emission from the oxygen-bearing species CO ($\nu$=\,1$\rightarrow$0) and \ce{CO2} (bending mode at 14.98\,$\mu$m). \cite{Arabhavi2025} reported weak detection of ro-vibrational water in this source.

The slab models provide valuable first estimates of the emitting conditions of the various molecules. However, such models do not account for dust opacity, chemistry or a self-consistent gas temperature given the emitting region around the star. 
The thermal structure, radiative transfer and the chemistry are intertwined. 2D thermo-chemical disk models possess a high degree of consistency in their physical and chemical structures. This implies that changing a parameter to increase the emission from one molecule also affects the emission from other molecules; this can have many reasons, e.g.\ the chemistry could be connected, the thermal balance of the gas could be affected or the dust optical depth changes and hence our ability to probe into deeper disk layers. So, contrary to individual 0D slab models, we can no longer tweak a single molecule at a time. Earlier studies such as \cite{Walsh2015, Greenwood2017, Kanwar2024} have modelled the disks around VLMS with 2D thermo-chemical models and noted the key chemical pathways leading to the formation of hydrocarbons as well as the relevance of dust evolution for the strength of the mid-IR lines emission.

The aim of this paper, is to investigate whether such 2D thermo-chemical disk models can produce at all the conditions conducive to the observed JWST-MIRI spectra of disks around VLMS. More specifically, we want to examine under which conditions such models generate the observed pseudo-continuum of \ce{C2H2}. We also aim to find an explanation for the dichotomy of the presence of a large variety of hydrocarbons along with the major oxygen carriers CO and \ce{CO2}. \ce{CO2} has also been found in a number of other VLMS disks \citep[e.g.,][]{Arabhavi2023, Kanwar2024, Arabhavi2025b}. 

The paper uses the JWST-MIRI/MRS spectrum of J160532 from \cite{Tabone2023} that has been reduced with version 1.8.4 of the JWST Science Calibration Pipeline and CRDS context jwst\textunderscore1017.pmap. Section\,\ref{Method} describes our approach in using 2D thermo-chemical disk models. We present a disk geometry that can explain the JWST-MIRI spectrum of J160532 in Sect.~\ref{Results} and discuss how and why we reached this specific model geometry in Sect.~\ref{Discussion}. Based on this, we propose how to break the degeneracy between oxygen depletion and carbon enrichment and provide an explanation for the co-existence of a rich hydrocarbon spectrum and \ce{CO2} emission. We discuss implications and limitations of our results in Sect.~\ref{dis} and present our main conclusions in Sect.~\ref{Conclusion}.

\section{Method}\label{Method}
To gain a deeper understanding of the physical and chemical structure of the disk and which characteristics can produce key features of the observed spectrum, we use the thermo-chemical protoplanetary disk model P{\tiny RO}D{\tiny I}M{\tiny O} \citep{Woitke2009, Woitke2016a, Kamp2017}. 

\subsection{Thermo-chemical modelling} \label{Thermo-chemical modelling}
P{\tiny RO}D{\tiny I}M{\tiny O} is used to simulate the disk around the VLMS J160532. The code self-consistently calculates the thermal, physical and chemical structure of the disk. It solves 2D continuum radiative transfer to calculate the dust temperature structure. It then calculates the chemistry and gas temperature structure, by balancing gas heating and cooling to determine the latter. We use the extended hydrocarbon chemical network developed by \cite{Kanwar2023} after expanding over the large DIANA network \citep{Kamp2017}. This gas-phase network can form species as large as \ce{C8H5+}. It forms ices (adsorption and desorption) for all the neutral species and radicals but does not consider surface chemistry. It also includes thermal decomposition and three-body reactions from the STAND network \citep{Rimmer2016}. We primarily use reactions and adsorption energies from the UMIST Rate12 database \citep{McElroy2013}. We solve the chemistry in steady state. The elemental abundances are adopted from \citet{Woitke2016a}. The model takes into account UV molecular shielding \cite[for details, see][]{Woitke2024}. The P{\tiny RO}D{\tiny I}M{\tiny O} version used here is 0e87fc6e.

\subsection{Line selection}

We use the fast line transfer FLiTs described in \cite{Woitke2018} to calculate the mid-infrared spectrum. This code performs the full continuum and line radiative transfer and takes into account line overlap and non-LTE effects for molecules such as CO and \ce{H2O}. We considered both ortho and para forms of \ce{H2O} while calculating its spectrum. No isotopologues of CO are considered as our disk chemistry does not account for isotopologue selective photodissociation.

FLiTs has the ability to capture both absorption and emission features of the spectrum. FLiTs produces two spectra: a dust continuum spectrum and a spectrum with dust continuum and molecular line emission on top. The dust continuum spectrum is subtracted from the total spectrum to obtain the molecular spectrum. The resulting spectrum is then convolved and resampled \citep[using spectres,][]{Carnall2017} to the JWST-MIRI/MRS wavelength grid. We use a resolution $R\,=\,3500$ for channel 1, 3000 for channel 2, and 2500 for channel 3. This makes the model output directly comparable to the manually continuum subtracted JWST-MIRI spectrum obtained from \cite{Tabone2023}.

We use the HITRAN2020 \citep{Gordon2022} spectroscopic database to obtain the Einstein coefficients, upper energy level $E_\mathrm{u}$, degeneracies and partition functions. The rules for line selection are generally adopted from \cite{Woitke2018} but modified for a few molecules as we observe more transitions with JWST-MIRI/MRS. The lines we select are between 4.89 to 20\,$\mu$m. All these lines are considered in the gas heating and cooling balance. The line strength is defined by Equation 2 in \cite{Woitke2018} and Table\,\ref{line_criteria} lists our selection rules for this work. We use the new selection criterion for \ce{C2H2} as this molecule shows a forest of weak lines that contribute to the formation of a pseudo-continuum. Hence, all available lines are considered. A total of 300 levels are considered for CO \citep{Thi2013} and we use collisional rates from \cite[H]{Balakrishnan2002}, \cite[He]{Cecchi2002, Krems2002}, \cite[\ce{H2}]{Yang2010} and \citet[\ce{e-}]{Ristic2007} to calculate the non-LTE spectra. Our selection consists of 60 rotational levels in the ground electronic state along with 5 vibrational levels in the ground state. A total of 6 and 26 bands are considered for \ce{CO2} and HCN, respectively. For water, we use molecular data from the Leiden LAMDA database \citep{vanderTak2020} including the collisional rates from \cite{Faure2008} to calculate non-LTE spectrum.

\begin{table}
    \centering
        \caption{Line Selection criteria for the HITRAN 2020 database.}
    \begin{tabular}{ccc} \hline
        Molecule & line strength [s$^{-1}$]  & $E_\mathrm{u}$ [K] \\ \hline
        \ce{C2H2} &  >\,10$^{-6}$ & <\,6000\\
        \ce{CH4} & >\,10$^{-3}$ & <\,4000\\
        CO & >\,10$^{-3}$ & <\,4000 \\ \hline
         & band selection & $\#$ of bands \\ \hline
        \ce{CO2} &`0 0 0 01'  `0 1 1 01'& 6 \\
                 &`0 1 1 01'   `0 2 2 01'& \\
                &`0 1 1 01'   `1 0 0 02'& \\
                &`0 1 1 01'   `1 0 0 01'& \\
                &`0 0 0 01'   `0 0 0 11'& \\
                &`0 1 1 01'  `0 1 1 11'& \\ \hline
        HCN &    `0 2 2 0' `0 1 1 0'  & 26 \\
             &    `0 1 1 0' `0 0 0 0' &  \\
             &    `0 3 1 0' `0 1 1 0' & \\
            &     `0 3 3 0' `0 2 2 0' & \\
            &     `0 4 2 0' `0 2 2 0' & \\
           &      `0 5 1 0' `0 3 1 0' & \\
             &       '0 4 4 0' '0 3 3 0' &  \\
       &   '0 3 1 0' '0 2 0 0' & \\
           &         '0 2 0 0' '0 0 0 0' &  \\
            &        '0 2 0 0' '0 1 1 0' & \\
             &       '0 4 0 0' '0 2 0 0' &  \\
              &      '0 5 3 0' '0 3 3 0' &  \\
               &      '0 6 2 0' '0 4 2 0' & \\
                &     '0 4 2 0' '0 3 1 0' &  \\
                 &     '0 5 5 0' '0 4 4 0'  & \\
 & `0 7 1 0' `0 5 1 0' & \\
 & `0 5 3 0' `0 4 2 0' &  \\
 &`0 6 4 0' `0 4 4 0' &  \\
 &`0 7 3 0' `0 5 3 0' &  \\
 &`0 3 1 0' `0 2 2 0' & \\
 &`0 4 0 0' `0 3 1 0' & \\
 &`0 6 0 0' `0 4 0 0' &  \\
 &`0 6 6 0' `0 5 5 0' &  \\
 &`0 5 1 0' `0 4 0 0' &  \\
 &`0 8 2 0' `0 6 2 0' &  \\
 &`0 6 4 0' `0 5 3 0' &  \\ \hline
    \end{tabular}
    \label{line_criteria}
\end{table}

\subsection{Disk structure}\label{Modelling strategy}

We did not perform a systematic exploration of all disk parameters as it is not feasible. This is because each model takes $\sim$11000\,seconds of CPU time. We applied observational constraints on parameters such as disk mass and stellar parameters. We explored a subset of parameters to arrive at a disk structure best describing the observations. We employ a disk model described by an inner and outer zone separated by a gap. 
This structure is similar to the one discussed in \cite{Tabone2023}, who also proposed a gap. We aim to deviate as little as possible from the canonical disk parameters of a T\,Tauri disk mentioned in \cite{Woitke2016a} for the outer disk. \cite{Woitke2016a}, \cite{Kurtovic2021}, and \cite{Pegues2021} demonstrate that the outer disks around T\,Tauri and VLMS are similar however, the size of these two types of disks can vary. Table\,\ref{Parameter} lists the parameters for both inner and outer disk. We assume the grain size in the model follows an MRN power law distribution \citep{Mathis1977} with a maximum grain size (a$_{\rm{max}}$) of 3\,mm. The outer disk has a minimum grain size (a$_{\rm{min}}$) of 0.05\,$\mu$m while the inner disk has 5\,$\mu$m. The relatively large grains in the inner disk are chosen to match the absence of the silicate feature in the JWST-MIRI/MRS observations. \citet{Kessler2007, Olofsson2009} show that grains larger than $\sim$5\,$\mu$m do not contribute to the silicate feature.   

We adjust the dust mass so that the photometric observations \citep{Barenfeld2016, Luhman2012} for the entire disk are roughly matched. We assume a gas-to-dust mass ratio of 1000 to set the initial gas mass of the disk. This is based on \cite{Greenwood2019} and \cite{Pinilla2013}, who demonstrated that for a viscosity of 10$^{-3}$, the disk beyond 10\,au can have a gas-to-dust mass ratio of 1000 after 1.5\,Myrs. The dust masses derived from SEDs are known to be degenerate \citep[e.g.,][]{Kaeufer2023}. Therefore, the uncertainty in the dust mass propagates to the gas mass. This is why we optimize it during our analysis. The final value of the disk mass is listed in Table\,\ref{Parameter} along with the other key model parameters. The dust mass is below the derived upper limit from the ALMA non-detection \citep{Barenfeld2016}. We assume 1\% of the total gas mass resides in the inner disk.

\begin{figure}[h]
   \centering
    \includegraphics[width=\linewidth]{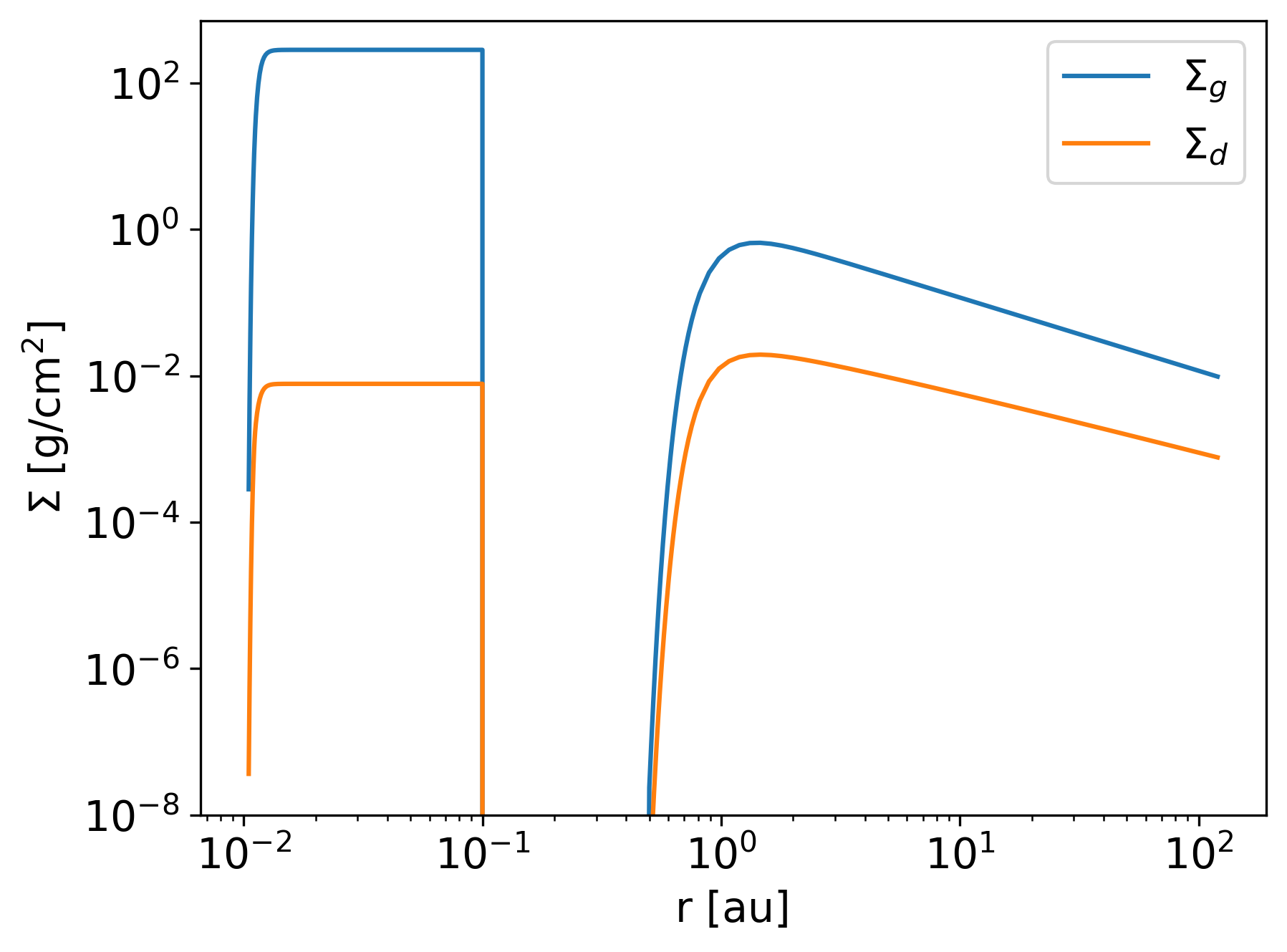}
   \caption{The full disk gas and dust surface density profile in the best model.}
              \label{gas_density}%
\end{figure}

The resulting surface gas and dust density profile for our best model is shown in Fig.\,\ref{gas_density}. We use a gradual build up of column density for the inner and outer disk as described in \cite{Woitke2024}. The column density power index $\epsilon$ for the inner disk is assumed to be zero, leading to a constant gas surface density profile which is the simplest assumption. The column density power index $\epsilon$ for the outer disk is assumed to be 1. As the emitting area of all the molecules derived using slab models in \cite{Tabone2023} is within 0.1\,au, we fix the outer radius of the inner disk to 0.1\,au so that the model is likely to reproduce the slab conditions that were retrieved from observations. We also varied other parameters in our quest for the best model, such as the gas-to-dust mass ratio, C/O elemental ratio in both zones, the location of the gap and the elemental abundances of O and C. A few of such models are listed in Appendix\,\ref{Different disk models}.

\begin{figure}[h]
    \centering    \includegraphics[width=\linewidth,keepaspectratio]{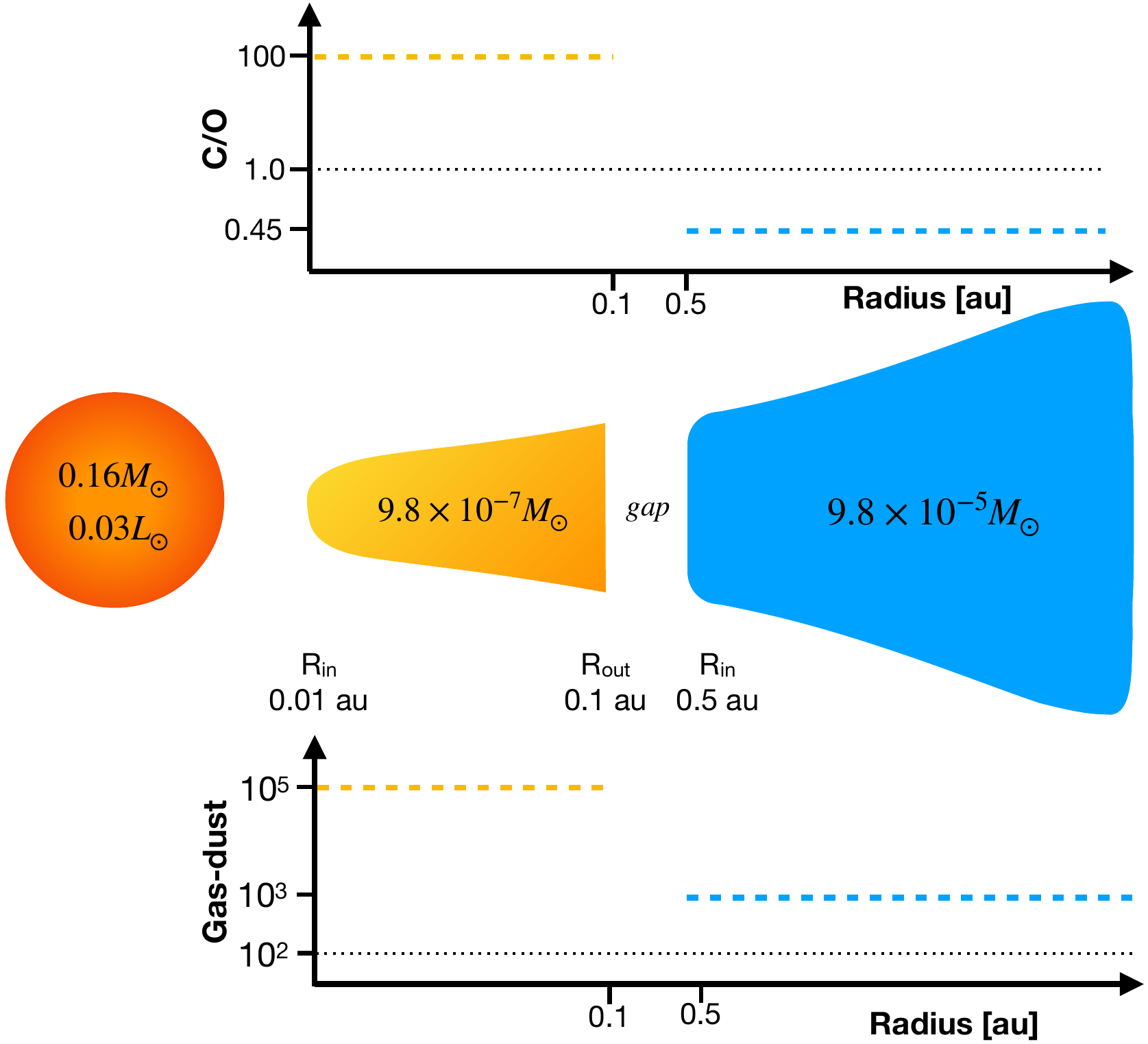}
    \caption{The illustration shows the best disk model along with the assumed C/O ratio and gas-to-dust mass ratio in the disk. The corresponding gas mass of the two zones are shown.}
    \label{im}
\end{figure}

The following list of molecules is used to explore a range of parameters with the 2D models to identify a disk model that captures the key observational aspects and provides a 2D context to the JWST spectrum: CO, \ce{CO2}, \ce{C2H2}, \ce{H2O}. We use these molecules as the spectrum is dominated by the pseudo-continuum of the hydrocarbon \ce{C2H2} and the other three are oxygen-bearing species, used to derive constraints also on the C/O ratio. We do not model the atomic and, molecular hydrogen lines and molecules such as \ce{C4H2} and \ce{C6H6} as they are beyond the scope of this study.




\subsection{Analysis}\label{Analysis}

 For each model, we used FLiTs to generate the spectrum of the above listed molecules, convolve and resample it to the JWST-MIRI/MRS wavelength grid and compare the observed flux levels with the modelled flux levels. The disk model where the peak line fluxes of molecules such as \ce{C2H2} in the range 13.54-13.75\,$\mu$m, \ce{CO2} in the range 14.96-14.98\,$\mu$m and most lines of CO are within a factor of 1.5 to the observed fluxes along with no observable water features is then considered to be the best model. It presents a possible disk geometry and elemental abundances that agree with the observations. A more detailed fitting of the observations with high accuracy is not the goal of this study.  

From the best disk model, we calculate the characteristics of the emitting region of these molecules such as the temperature $\langle T \rangle$, column density $\langle \log_{\rm {10}} N\rangle$ and the emitting area $\langle r_{\rm{1}} - r_{\rm{2}} \rangle$. They are calculated using the escape probability method assuming a face-on disk \citep{Woitke2024} and are statistical averages over the individual line emitting regions of all the lines in a given spectral range and over all the vertical columns of a selected molecule. The total integrated flux generated by each column and the line of that molecule is chosen as its statistical weights.

\begin{table}[]
\caption{Parameters for the model that best explains the observed fluxes.}
\label{Parameter}
\resizebox{\linewidth}{!}{%
\begin{tabular}{llll}
\hline
Quantity                            & Symbol                          & Values                            & Comments \\ \cline{1-4}
\multicolumn{2}{l}{Stellar parameters}                                &                                    \\ \cline{1-4}
stellar mass                        & $M_\star$                        & 0.16\,$M_{\sun}$ \tablefootmark{a}   &                \\
stellar luminosity                  & $L_{\star}$                     & 0.03\,$L_{\sun}$ \tablefootmark{b}     &               \\
effective temperature               & $T_{\rm{eff}}$                       & 3100\,K                           & \\
UV excess$^*$                           & $f_{\mathrm{UV}}$               & 0.026                              &\\
UV powerlaw index                   & $p_{\mathrm{UV}}$               & 0                                  &\\ 
Mass accretion rate                 & $\dot{M}_{\rm acc}$                & 10$^{-9}$\,$M_{\odot}yr^{-1}$ & \\
X-ray luminosity                   & $L_X$                & 6$\times$10$^{28}$\,erg\,s$^{-1}$ &\\
X-ray emission temperature         & $T_X$                & 2$\times$10$^{7}$\,K &\\ \cline{1-4}
distance                            & $d$                               & 152\,pc                            &\\
inclination                         & $i$                               & 45\textdegree                    &\\ 
grid size                           & radial$\times$vertical          & 200$\times$150                   & \\ \hline
\multicolumn{3}{l}{Dust properties}                                                                        \\ \hline
maximum dust particle radius        & $a_\mathrm{max}$                & 3000\,$\mu$m                        &\\
settling method                     & settle\textunderscore method                   & Riols\,\&\,Lesur     &                 \\
settling parameter                  & $a_\mathrm{settle}$ or $\alpha$ & 10$^{-3}$                         & \\ \hline
strength of interstellar UV         & $\chi^\mathrm{ISM}$             & 1                                 & \\
cosmic ray \ce{H2} ionization rate  & $\zeta$                         & 1.3\,$\times$\,10$^{-17}$ s$^{-1}$ &\\ \hline
\multicolumn{3}{l}{Outer disk parameters}                                                                  \\ \hline
minimum dust particle radius        & $a_\mathrm{min}$                & 0.05\,$\mu$m                         &  \\
disk gas mass$^*$                       & $M_{\mathrm{disk}}$                      & 9.85$\times$10$^{-5}$\,$M_{\sun}$  &  10$^{-6}$-10$^{-4}\,M_{\sun}$\\
dust-to-gas ratio$^*$                   & dust-to-gas ratio               & 0.001                               & 10$^{-2}$-10$^{-3}$\\
inner radius of outer disk$^*$ & $R_{\mathrm{in}}$                        & 0.5\,au                             & 0.4 to 10\,au\\
outer disk radius                   & $R_{\mathrm{out}}$                       & 120\,au                          &   \\
carbon-to-oxygen ratio$^*$              & C/O                       & 0.45                               & 0.01 to 100\\
column density power index          & $\epsilon$                      & 1                                 & \\
reference scale height    & $H_\mathrm{g}(100\,{\rm au})$ & 5\,au &\\
extension                           & raduc1                          & 2.00                               &\\
maximum $\Sigma$ reduction          & reduc1                          & 10$^{-8}$                        &  \\
flaring index                       & $\beta1$                         & 1.15                              &  \\ \hline
\multicolumn{3}{l}{Inner disk parameters}                                                                  \\ \hline
minimum dust particle radius        & $a_\mathrm{min}$                & 5.00\,$\mu$m                          & \\
disk gas mass$^*$                       & $M_{\mathrm{disk}}$                      & 9.85$\times$10$^{-7}$\,$M_{\sun}$   & 10$^{-8}$-10$^{-6}\,M_{\sun}$\\
dust-to-gas ratio$^*$                   & dust-to-gas ratio               & 1.33$\times$10$^{-5}$               & 10$^{-2}$-10$^{-6}$\\
inner disk radius                   & $R_{\mathrm{in}}$                        & 0.0105\,au                          &\\
outer radius of the inner disk$^*$      & $R_{\mathrm{out}}$                       & 0.1\,au                           & \\
carbon-to-oxygen ratio$^*$              & C/O                       & 100                                & 0.45 to 100\\
column density power index          & $\epsilon$                      & 0                                 & \\
reference scale height    & $H_\mathrm{g}(0.3\,{\rm au})$ & 0.003\,au & \\
extension of inner rim$^*$              & raduc2                          & 1.10                              & \\
maximum $\sigma$ reduction          & reduc2                          & 10$^{-6}$                          &\\
flaring index                       & $\beta2$                         & 1.15 &\\
\hline
\end{tabular}}
\tablefoot{The free parameters are indicated by *. The range in which certain parameters were changed are stated in the last column. \tablefoottext{a}{\cite{Pascucci2013}} \tablefoottext{b}{\cite{Carpenter2014}}}
\end{table}

\section{A 2D thermo-chemical model for J160532}\label{Results}

We present our identified best disk geometry in Fig.\,\ref{im} with the parameters listed in Table\,\ref{Parameter}. We first describe the key results from this model and then provide the arguments driving this choice of parameters in Sect.~\ref{Discussion}.

\begin{figure*}[ht!]
   \centering
    \includegraphics[width=\linewidth]{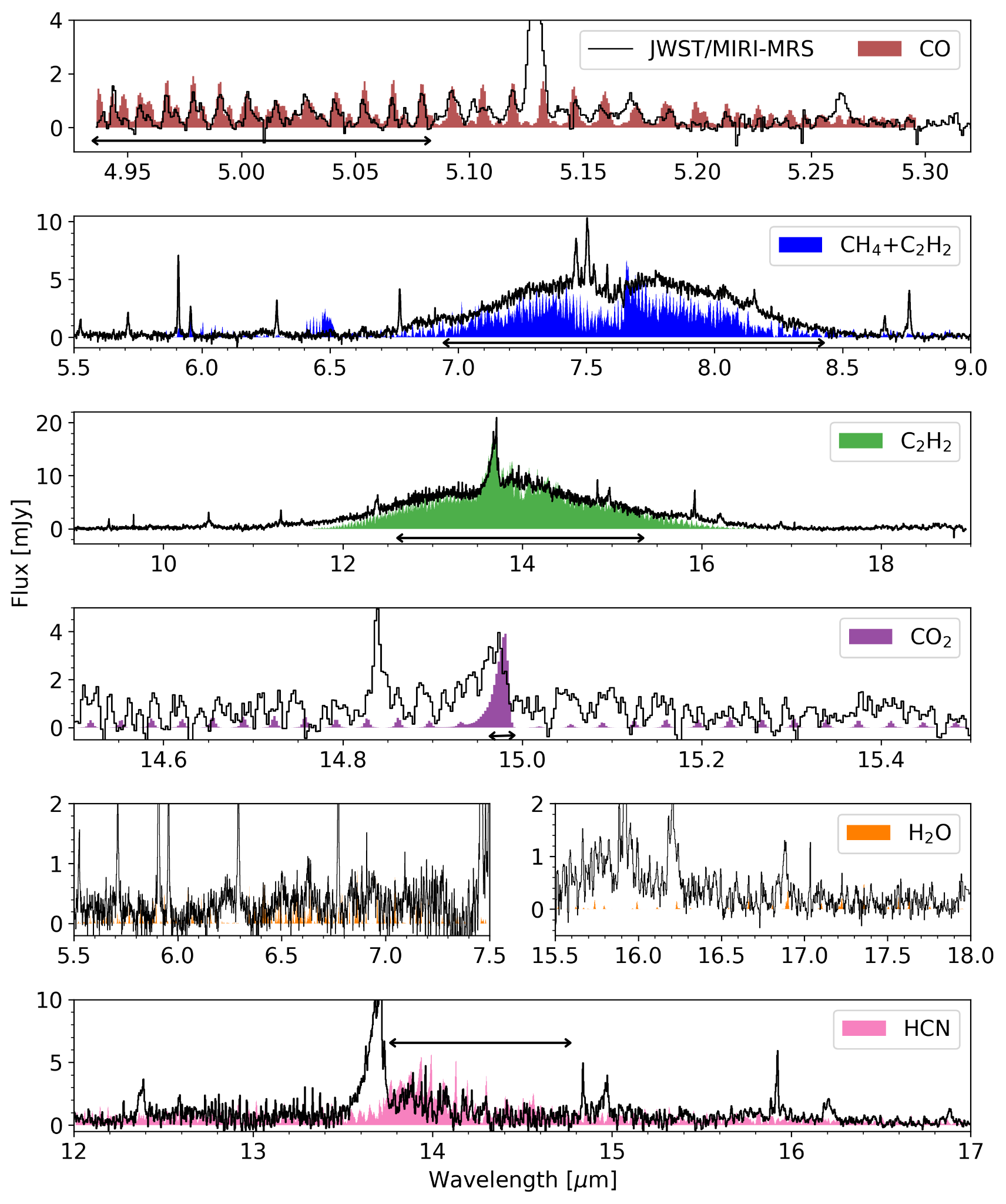}
   \caption{The modelled spectrum of the best disk model convolved and resampled to the JWST-MIRI/MRS resolution and compared to the continuum subtracted observed spectrum. \citet[Extended Data Figure\,5]{Tabone2023} provide two different continua and therefore, we use the continuum from their middle panel to depict \ce{C2H2} and \ce{CH4} whereas the continuum from their bottom panel is used to depict the rest of the molecules. The solid black line depicts the region where the modeled fluxes approximately match the observations. These are the regions where little effect from blending is expected.}
              \label{spectrum}
\end{figure*}

The molecular emission from our best model convolved and resampled to the JWST-MIRI/MRS resolution matches well with the observed spectrum (Fig.\,\ref{spectrum}). As mentioned in Sect.\,\ref{Modelling strategy}, the disk is divided into an inner and outer disk separated by a gap. The outer disk resembles a typical disk with a 10 times increase in gas-to-dust mass ratio relative to the canonical value of 100. The inner disk is extremely depleted in elemental oxygen and dust. The gas and dust thermal structure along with the gas density and UV radiation field are shown in Fig.\,\ref{properties}. The disk is largely optically thin, with the maximum optical depth of 0.7 acquired at 1.5\,au at UV wavelengths ($\lambda\!=\!0.096\,\mu$m). This is a result of the presence of only large ($>5.0\,\mu$m) grains in the inner disk. This best model is able to produce the molecular pseudo-continuum of \ce{C2H2}. The model shows the large column densities of \ce{C2H2} required to obtain it. In our model, this is possible because of the low dust opacity. \cite{Woitke2024} also found that a low continuum optical depth exposed the \ce{C2H2} molecular reservoir and produced a strong emission feature. 

The peak flux of \ce{C2H2} matches the JWST observations whereas the P-branch is under predicted by $\sim$28\% relative to the observations and the R-branch is over-predicted by $\sim$24\% (Fig.~\ref{spectrum}). The modelled flux of CO is at most $\sim$86\% higher relative to the observations. The modelled \ce{CO2} peak emission is lower in flux relative to the observations by $\sim$33\%. 

The lack of emission from our model at $\sim\,$14$\,\mu$m in the third panel of Fig.\,\ref{spectrum} is explained by the presence of HCN emission in the observations. When calculating the combined spectrum of \ce{C2H2} and HCN with FLiTs taking into account the line opacities and overlap, we obtain flux levels close to the observations (within 10\%, see Figure\,\ref{C2H2+HCN}). This confirms that HCN is blended with \ce{C2H2} as shown in \cite{Tabone2023}.

The model underpredicts \ce{C2H2} at short wavelengths (6.8-8.5\,$\mu$m) despite the contribution of emission from \ce{CH4}. At these short wavelengths, we are limited by the incomplete spectroscopic line data for \ce{C2H2} in HITRAN. \cite{Tabone2023} proposed the presence of excited $^{13}$CCH$_2$; however, we do not consider $^{13}$CCH$_2$ in our model as we are again limited by its incomplete spectroscopic data. In addition, the model reproduces the observations at long wavelengths (11-16\,$\mu$m) fairly well using only the main isotopologue (Fig.~\ref{spectrum}, third panel). 

To demonstrate the effect of the lack of spectroscopic data on the \ce{C2H2} molecular fluxes, we select only the $\nu_5$ band of \ce{C2H2}, the strongest mode which dominates the entire spectrum. We then re-calculate the molecular emission of \ce{C2H2} using FLiTs from the best-fit model (keeping the gas temperature fixed) and find that we under-predict the peak flux of \ce{C2H2} at 13.71\,$\mu$m by a factor of $\sim$5. Thus, weak lines are crucial in producing the molecular pseudo-continuum and this exercise illustrates the need for a more complete spectroscopic database. 

As a sanity check, we also confirm that our best disk model produces only weak water features at short or long wavelengths (Fig.~\ref{spectrum}, fifth panel).

\subsection{Model-based interpretation of the JWST spectrum} \label{Modelled spectrum}



\begin{figure}
   \centering
    \includegraphics[width=\linewidth]{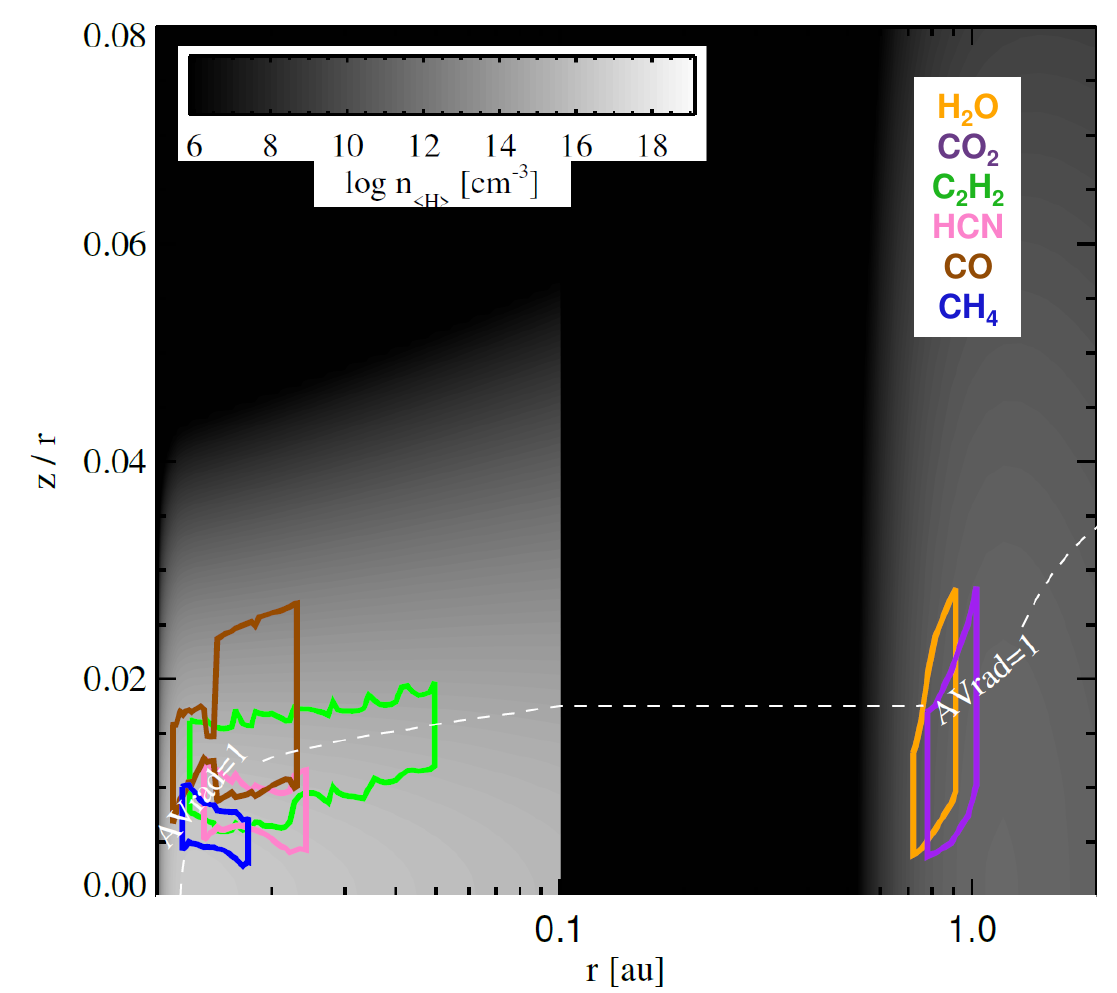}
   \caption{The emitting region (4.89-19\,$\mu$m) of different molecules in the model. The dashed contour corresponds to the radial A$_{\rm{v}}$ of 1\,mag. The disk model is optically thin in the vertical direction.}
              \label{EA}%
\end{figure}

The emitting areas of our key molecules are shown in Fig.\,\ref{EA} for the best model. A complete overview of 2D molecular abundances can be found in Fig.\,\ref{abundances} in the appendix. The molecular emission of molecules such as \ce{C2H2}, \ce{CH4}, \ce{HCN}, \ce{CO} originates from the inner disk. However, \ce{CO2} and faint \ce{H2O} emission originate behind the inner wall of the outer disk. This is in-line with results from \cite{Vlasblom2024}, who also find \ce{CO2} to be excited in the cooler gas beyond disk gaps. The molecular emission of \ce{CH4} is coming from deeper layers relative to CO and \ce{C2H2} which are somewhat co-spatial. However, the emitting region of \ce{C2H2} is radially more extended than \ce{CO}. There is a tenuous surface layer of \ce{H2O} in the inner disk, but it produces no significant emission. Most emission of \ce{H2O} in our best model is originating from the outer disk; however the gas here is not warm enough to produce strong features. 

To quantify the emitting conditions of the various molecules in our best 2D thermo-chemical disk model, Table\,\ref{TNR} shows the values of gas temperature $\langle T \rangle$, column density $\langle \log_{\rm {10}} N\rangle$ and the radial extent of the emitting region $\langle r_{\rm1} - r_{\rm 2} \rangle$ derived using the line fluxes based on the escape probability method. The ranges shown are the standard deviation of the statistical average from all lines inside the wavelength range 4.9-19$\,\mu$m. 

We cannot compare the values obtained from the best disk model in Table\,\ref{TNR} directly to those of the slab models \citep{Tabone2023} as the disk model shows abundance and temperature gradients which are not always well-captured in the averages obtained in Table~\ref{TNR} \citep{Kamp2023}. \cite{Kaeufer2024} demonstrated that the JWST spectra of certain molecules are indeed better fit by radial power law slab models.

We find that the weak water emission in the JWST spectrum still implies $\langle \log_{\rm {10}} N\rangle$ of $\sim\,$16 at $\langle T \rangle$ of $\sim\,$300\,K in the presence of a C/O ratio of 100. The $\langle \log_{\rm {10}} N\rangle$ obtained from disk model for \ce{H2O} is well-below from what is found in \cite{Tabone2023}. Note we find larger emitting area and a slightly different $\langle T \rangle$. We find \ce{CO2} temperatures cooler than those derived in \citet{Tabone2023}. In our model, the temperature range for both \ce{H2O} and \ce{CO2} is low as they emit from the outer disk over a small temperature gradient. Hence, for \ce{CO2} we find much larger emitting area and cooler temperature contrary to \cite{Tabone2023}. They reported \ce{C2H2} is warmer than \ce{CO2}. This trend is similar to what we find. 

CO is the hottest molecule in the best model ($\sim\,1400$~K) emitting with high column densities ($\langle \log_{\rm {10}} N\rangle\,\sim\,20$), albeit from very close in (0.02\,au). Despite the decrease of oxygen by a factor $\sim\,100$, we find high levels of emission from CO. This also shows that CO gets optically thick easily. The slab model retrievals for temperature of CO are cooler for VLMSs than T\,Tauri stars hinting that the disks around VLMSs are cooler \citep{Pascucci2013}. However, we find hot temperatures for CO and also detect the $\nu$=2$\rightarrow$1 transition indicative of warm inner regions.
\ce{CH4} also has high column densities ($\langle \log_{\rm {10}} N\rangle\,\sim\,21.5$), high temperatures ($\langle T \rangle\,\sim\,1150$~K) and a similar emitting area. This produces a clear emission feature at 6.5$\mu$m (see Fig.\,\ref{spectrum}, second panel). It is the second hottest molecule in our model and it is also emitting from a very narrow region closer to the central star. 

We find high column densities for \ce{C2H2} similar to the findings of \cite{Tabone2023} and the formation of a pseudo-continuum. As the emitting area of \ce{C2H2} is large ($\sim\,$0.05\,au) and vertically extended, the temperature gradient in this emitting region produces the large range of gas temperatures ($\langle T \rangle\,\sim\,818\pm503$~K). The values inferred from 0D slab models, $T=400-525$~K lie well within the range of temperatures inferred from our 2D model. 
The 2D structure automatically comprises optically thick and thin emitting conditions within the reported emitting region and has thus no problem to reproduce the full band shape. We will come back to this also in Sect.~\ref{sect:C2H2}. 

The emission of HCN is blended with \ce{C2H2}. \cite{Tabone2023} provide an upper limit on column density of HCN after fixing the temperature and the emitting area. We quantify the emitting conditions of HCN using our self-consistent model (see Table\,\ref{TNR}). We find large column densities in a smaller emitting region relative to the previously reported values from the 0D slab models. Overall, the spread in the values obtained from the 2D disk model does include the values retrieved from the 0D slab models of \citet{Tabone2023}. 
\begin{figure}[h]
   \centering
    \includegraphics[width=\linewidth]{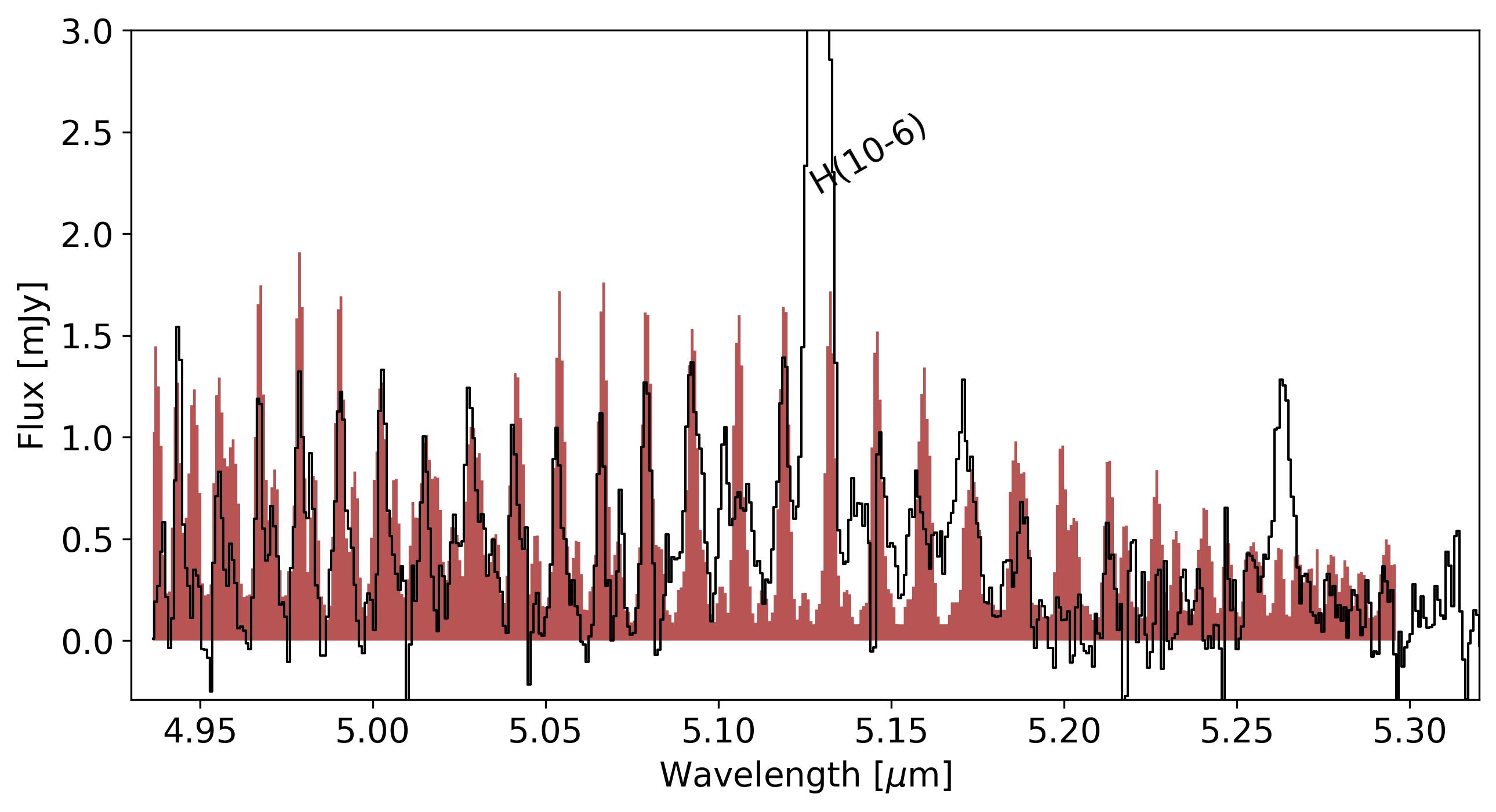}
   \caption{The zoom-in JWST/MIRI MRS spectrum (black) showing CO emission. The emission of CO from the best thermo-chemical model convolved to 3500 and resampled to JWST-MIRI/MRS resolution is shown in brown. The H\,{\sc i} 10-6 line is marked in grey.}
              \label{CO}%
\end{figure}

\subsection{Detection of CO $\nu$= 2$\rightarrow$1}

\cite{Tabone2023} report the detection of the CO $v=1\rightarrow0$ fundamental band. Triggered by the spectrum derived from our best disk model, we report here also the detection of the CO $v=2\rightarrow1$ band. CO can either be excited by UV \citep{Krotkov1980}, IR pumping \citep{Scoville1980} or by collisions \citep{Bosman2019}. Figure~\ref{CO} zooms in on the modelled CO emission along with the JWST/MIRI-MRS observation. The integrated flux predicted from the model matches the observed emission within 50\%. 
The full shape of the CO ro-vib band at 4.7\,$\mu$m (see Fig.~\ref{CO_nirspec} for the full NIRSpec range using $R\,=\,2700$) is an intricate superposition of the $v=1\rightarrow0$ and $v=2\rightarrow1$ bands, showing for example a minimum at 5.02\,$\mu$m, clearly visible both in the model and the observation.

\begin{figure}[h]
   \centering
    \includegraphics[width=\linewidth]{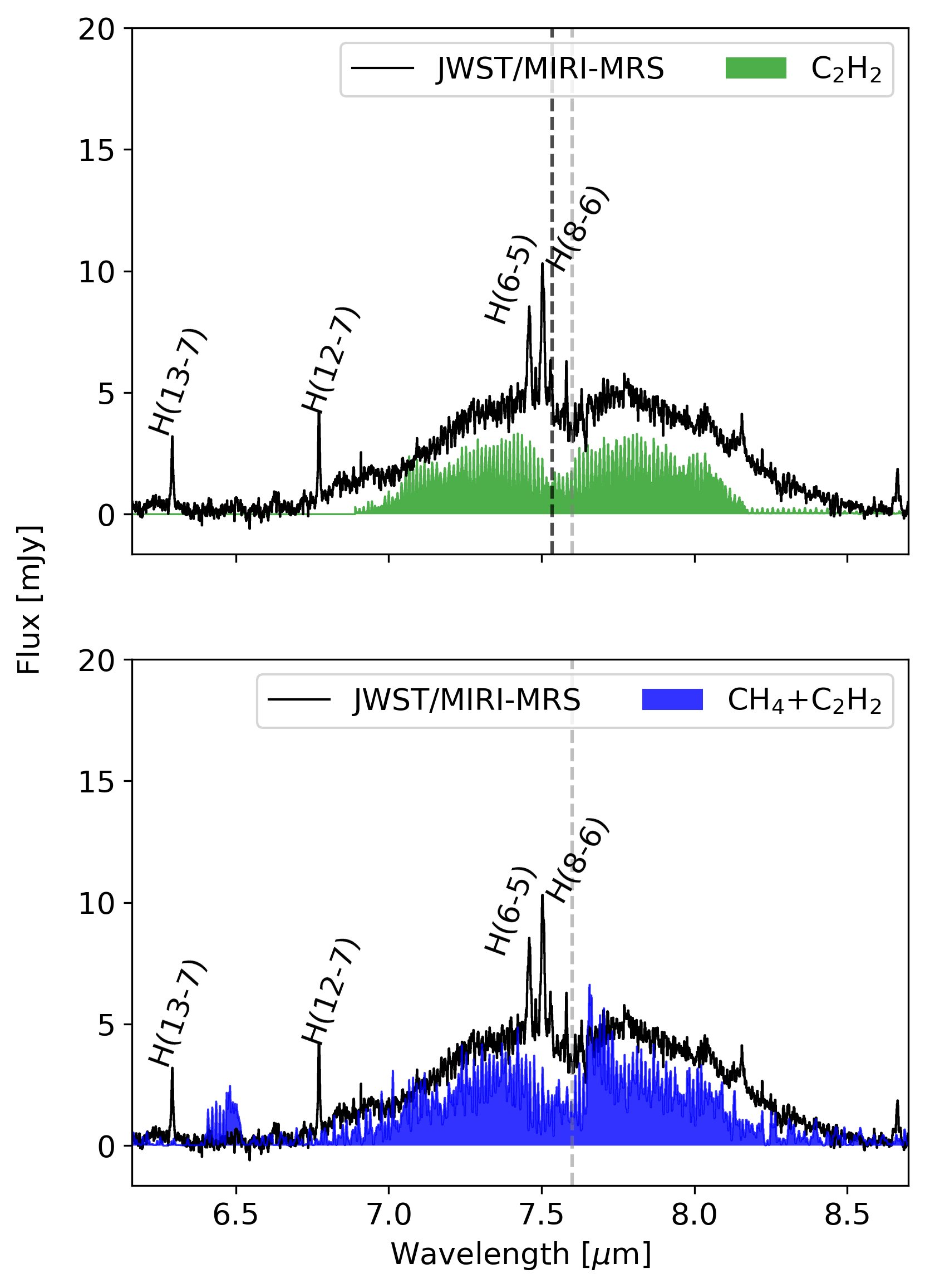}
   \caption{The top panel shows the modelled flux of \ce{C2H2} and the bottom panel shows the modelled flux of \ce{C2H2} and \ce{CH4} together using FLiTs. The gray dashed line depicts where the 'dip' in observed spectrum occurs. The black dashed line shows the dip in flux in the best modelled spectrum.}
              \label{C2H2+CH4_diff}%
\end{figure}

\subsection{\ce{CH4}}

\cite{Tabone2023} reported a tentative detection of \ce{CH4}. Here we present further evidence of its presence in the spectrum, albeit being strongly blended with \ce{C2H2} molecular emission. Figure\,\ref{C2H2+CH4_diff}, shows the modelled molecular emission of \ce{C2H2} only in the top panel and the combined emission of \ce{C2H2} and \ce{CH4} in the bottom panel including line blending and opacity overlap of both molecules. The emission from the P- and R-branch of \ce{C2H2} is symmetric about the black dashed line in the modelled spectrum (see Fig.~\ref{C2H2+CH4_diff}, top panel). However, in the observations, the minimum emission in the butterfly pattern occurs at a location that is red-shifted (gray dashed line) compared to the model. This shift can be explained by the presence of the triply degenerate deformation $\nu_4$ mode of \ce{CH4} at 7.65\,$\mu$m. The combined spectrum of \ce{C2H2} and \ce{CH4} in the bottom panel aligns well with the observations and the emission minimum in the modelled spectrum and the observations occurs at the wavelength indicated by the gray dashed line (see Fig.~\ref{C2H2+CH4_diff}, bottom panel). The emission at 6.5\,$\mu$m corresponds to the doubly degenerate deformation $\nu_2$ mode of \ce{CH4}, which seems slightly overpredicted by our model. Still, we conclude that the broad emission feature at 7.5\,$\mu$m can be explained best by the presence of both \ce{CH4} and \ce{C2H2}.  

\ce{CH4} could be overpredicted because either it is less abundant than \ce{C2H2} thus lowering $\langle \log_{\rm {10}} N\rangle$ or it is much cooler than our model predicts. As it is a self-consistent model, unlike slab models, the abundance is given by chemical considerations and is not a free parameter. This difference in abundance could be due to uncertainties in the rate coefficients of the reactions which affect the final abundances of the species.

Figures~\ref{CO} and \ref{C2H2+CH4_diff} (bottom panel), show that the continuum baseline could also be placed a bit lower, thus bringing our model predictions for CO at $\sim$4.98\,$\mu$m and \ce{CH4} at 6.5\,$\mu$m in better agreement with the observations.


\begin{table*}
    \centering
    \caption{The characteristics of the emitting region of various molecules in the best disk model.}
    \begin{tabular}{ccccccccc}
    \hline 
    
Molecule & \#lines &  $\Sigma$ $F_{\rm line}$\,[Wm$^{-2}$]&  $\langle T \rangle$\,[K]&  $\langle$ log$_{\rm {10}} N\rangle$\,[cm$^{\rm {-2}}$]& $\langle r_{\rm1} - r_{\rm 2} \rangle$\,[au] & T\,[K] & log$_{\rm {10}}$ N\,[cm$^{\rm {-2}}$]& r\,[au] \\ \hline 
CO       &  181  &  3.6(-17)&  1411$\pm$ 640 &  20.0$\pm$ 0.5& 0.01-0.02 & & &\\
\ce{CH4}& 3353 &  1.3(-16)& 1145$\pm$324 & 21.4$\pm$0.8 & 0.01-0.02 & & & \\
\multirow{2}{*}{\ce{C2H2}}&  \multirow{2}{*}{24450} & \multirow{2}{*}{7.4(-16)} & \multirow{2}{*}{818$\pm$503}  & \multirow{2}{*}{20.4$\pm$0.2} & \multirow{2}{*}{0.01-0.05} & 525 & 20.38 & 0.03\\ & & & & & & 400 & 17.39 & 0.07 \\
\ce{H2O} &  3215 &  7.7(-18)&  301$\pm$ 66   &  16.1$\pm$ 0.5& 0.72-0.97 & 400 & <17.9 & 0.07\\
\ce{CO2}&    838  & 1.4(-18)  & 275$\pm$24  & 14.7$\pm$0.4& 0.8-1.07 & 430 & 18.37& 0.03\\
\ce{HCN}&  5673 & 1.8(-16) & 912 $\pm$ 283 & 20.2 $\pm$ 0.8 &0.01-0.03 & 400 &17.17 & 0.07\\ \hline
    \end{tabular}
    \label{TNR}
    \tablefoot{The average properties are calculated within the spectral region of 4.9-19\,$\mu$m. The range given is the standard deviation from the statistical average. The second column states the molecular flux as A$\times$10$^{-B}$ represented as A(-B). The last three columns list the parameters retrieved from observations via 0D slab models from \cite{Tabone2023}. The parameters for both the optically thick and thin component of \ce{C2H2} derived from the observations are reported.}
\end{table*}


\section{Justifying the best model}\label{Discussion}

In the following we present the lessons learned from exploring a larger parameter space to arrive at our best model. We will provide the justification for deciding on parameters such as the gap location, dust-to-gas mass ratio, element abundances, and suggest diagnostics to be developed further for a more general interpretation of JWST spectra of disks.

\subsection{Why and where to put a gap}

When modelling a disk without any gap and using a high gas-to-dust mass ratio (10$^{5}$) and high C/O ratio (100), we are not able to simultaneously produce the molecular fluxes of all the molecules we considered in this analysis. Increasing the gaseous C/O ratio in a continuous disk resulted in extremely weak molecular emission of \ce{CO2} and strong emission of CO even at a C/O ratio of 100 (see Fig.\,\ref{CO2_fulldisk}). Hence, we introduced a gap in the disk, leading to an inner and outer disk. 

The outer disk is acting as an oxygen reservoir. For example, if both the inner and outer disks have a high C/O ratio (inner: C/O\,=\,100, outer: C/O\,=\,10) , we still lack detectable emission from \ce{CO2} (see Fig.\,\ref{CO2_gapdisk}). Hence, an oxygen reservoir in the outer disk is required to produce detectable \ce{CO2} molecular emission, unless there is any other efficient mechanism to produce \ce{CO2} when the C/O ratio is high. If we use a low C/O ratio in the outer disk, CO closely matches the observation, because its emission originates from the inner disk with the high C/O ratio (C/O\,=\,100). Therefore, these C/O ratios are adopted in the best model (Sect~\ref{Results}).

We are only able to reproduce the spectrum with the given parameter space exploration when the disk has a gap present from 0.1 to 0.5\,au. The flux levels of \ce{CO2} and \ce{CO} are very sensitive to the location of this gap. Although, the best-fit model does not perfectly match the observed line fluxes, moving the inner edge of the outer disk inwards, increases the line flux for \ce{CO}. This is because the CO in the outer disk becomes hot and starts to emit from the outer oxygen-rich disk. Thus, even if the fluxes only match within a factor of at most 1.5, the location of the outer edge of the gap is well-constrained within our modelling framework. Moving the inner edge of the outer disk outwards decreases the \ce{CO2} flux as the gas becomes too cool to produce detectable features in the mid-IR spectrum. If the outer edge of the inner disk is moved outwards, the emitting area of species such as \ce{C2H2} increases along with the decrease in the column density, thus again deviating from the observations.

\subsection{The dust-to-gas ratio in the inner disk} 
\label{subsect:d2g}

The inferred molecular column densities of \ce{C2H2} over the dust continuum are huge, $\sim\,$10$^{20}$\,cm$^{-2}$ \citep{Tabone2023}. Such column densities can only be achieved in our model by depleting the disk in dust and lowering the dust-to-gas mass ratio to 10$^{-5}$. This makes the inner disk almost optically thin (see Fig.~\ref{EA}). An alternative way to achieve a high column density for \ce{C2H2} is by increasing the gas mass of the inner disk. This could help to reduce the C/O ratio required to obtain the high \ce{C2H2} column densities. However, an increase in gas mass will also lead to a rise in CO flux. To counteract this, the C/O ratio must be even further increased beyond 100. Therefore, the presence of both the pseudo-continuum of \ce{C2H2} and \ce{CO} can be used to better constrain the C/O ratio in the inner disk.

The dust-to-gas ratio is assumed 0.001 in the outer disk. If we use a dust-to-gas ratio of 0.01 instead, only the peak fluxes of \ce{CO2} and \ce{H2O} subtly decrease, because only these two molecules are emitting from the outer disk.



\subsection{Depletion of oxygen}

\begin{figure}[h]
   \centering
    \includegraphics[width=0.85\linewidth]{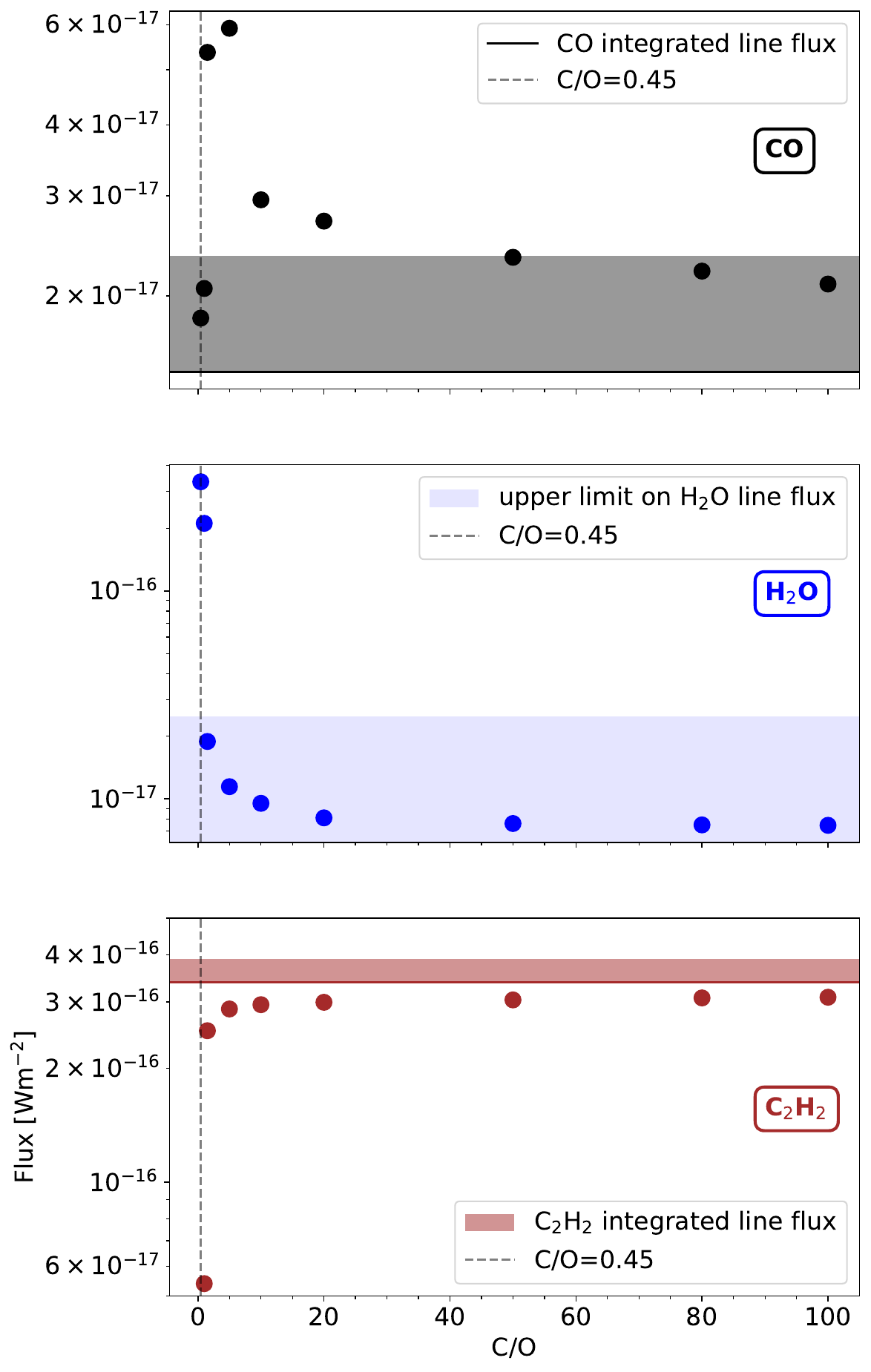}
   \caption{The effect of depletion of oxygen to enhance the C/O ratio on the integrated line flux of CO (4.936-5.298\,$\mu$m), \ce{H2O} (4.891-18.474\,$\mu$m) and \ce{C2H2} (11.664-17.289\,$\mu$m). The black horizontal line shows the CO observed integrated flux with gray area indicating the error, the blue shaded area shows the \ce{H2O} upper limit, and the brown line and shaded area the integrated observed flux of \ce{C2H2} and its associated error. The \ce{C2H2} flux for the model with a C/O ratio of 0.45 was low ($\sim$6$\times$10$^{-19}$\,Wm$^{-2}$) and is therefore not shown. The dotted vertical line highlights the model with a typical C/O ratio of 0.45.} 
              \label{C/O_ratio}%
\end{figure}

One can change the C/O ratio in the disk by depleting elemental oxygen or enriching the disk in carbon. 
We vary the C/O ratio using values of 0.45, 1, 1.5, 5, 10, 20, 50, 80, 100 in the disk by depleting
the elemental oxygen. Figure\,\ref{C/O_ratio} shows the effect of this depletion on the integrated line flux of CO between 4.93-5.29\,$\mu$m, \ce{H2O} between 4.89-18.47\,$\mu$m and \ce{C2H2} between 11.66-17.28\,$\mu$m. The black solid line indicates the integrated observed CO flux in the respective wavelength range and the uncertainty is calculated on the basis of the continuum placement. To calculate this uncertainty, we placed the continuum at the base of the line features; some of the CO lines drop below the continuum baseline assumed in \cite{Tabone2023} (see Fig.~\ref{CO}). Placing a more conservative continuum provides an upper limit on the integrated CO flux. The \ce{H2O} upper limit is calculated by integrating the flux produced by a 0D slab model with 400\,K, a column density of 8$\times$10$^{17}$ cm$^{-2}$ and an emitting area of 0.07\,au over 4.89-18.47\,$\mu$m. We integrate only the \ce{C2H2} band at the longer wavelength for comparison, because we know that we underpredict the flux at  short wavelength due to lacking molecular data (see Sect.~\,\ref{Modelled spectrum}).

An increase in the C/O ratio leads to a decrease in line fluxes of CO and \ce{H2O}. Oxygen is the limiting factor here in the formation of CO and \ce{H2O}. Thus, the abundance of CO decreases with depletion in oxygen. We get closer to the observed value with a C/O of 100 (see top panel of Fig.~\ref{spectrum} and Fig.~\ref{C/O_ratio}). Interestingly, we also notice an increase in CO flux between a C/O of 0.45 and 5. This is due to a gas temperature increase that results from lower \ce{H2O} line cooling. $^{12}$CO gets optically thick fast as we traverse deeper into the disk. The models show that the gas keeps getting hotter leading to the production of strong CO flux levels even when the C/O ratio is as large as 100. This also explains its shallow dependence on the C/O ratio for values larger than 50. Noticeably, CO and \ce{C2H2} can co-exist even when the C/O ratio is large.

The \ce{C2H2} flux rises drastically once the C/O ratio reaches a value of 1.5, and then keeps gradually rising with increasing C/O ratio (see bottom panel of Figs.\,\ref{C/O_ratio} and \ref{fUV}). At the C/O ratio of 5, the model is able to form a strong molecular pseudo-continuum given the low dust-to-gas mass ratio (Sect.~\ref{subsect:d2g} and Fig.\,\ref{fUV}). However, the C/O ratio of 5 still over-predicts CO and we require a further decrease of the oxygen abundance. The modelled emission better matches the observations when the C/O ratio is 100. Thus, in our modeling framework, a low C/O ratio is not able to produce the required molecular pseudo-continuum simultaneously with a low CO line flux. The presence of a molecular pseudo-continuum of \ce{C2H2} and low flux levels of \ce{CO} are a powerful diagnostic to better constrain the C/O ratio in both the inner and outer disks. 

\subsection{Enhancement of carbon}

\begin{figure}[h]
   \centering
    \includegraphics[width=0.88\linewidth]{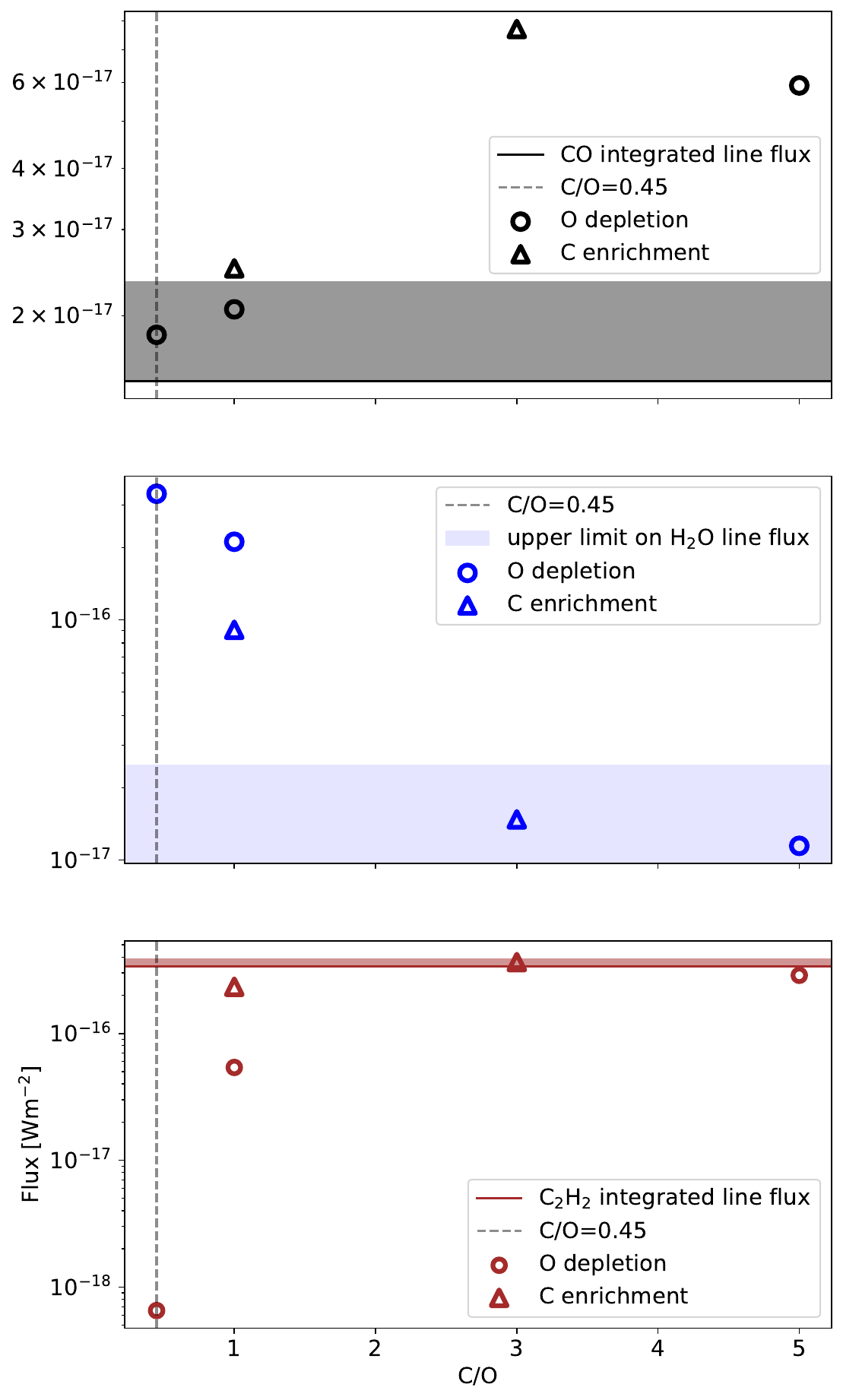}
   \caption{The effect of depletion of oxygen and enrichment of carbon to enhance the C/O ratio on the integrated line flux of CO (4.936-5.298\,$\mu$m) and \ce{H2O} (4.891-18.474\,$\mu$m) and \ce{C2H2} (11.664-17.289\,$\mu$m). The circles and triangles represent models where oxygen was depleted and carbon was enriched, respectively, to attain a high C/O elemental ratio. The solid horizontal brown line depicts the integrated \ce{C2H2} flux. The solid black line denotes the integrated observed line flux of CO and the dotted vertical line shows the model with a typical C/O ratio of 0.45 similar to Fig\,\ref{C/O_ratio}.}
              \label{C/O_ratio_enrichC}%
\end{figure}

When enriching the inner disk with carbon leading to a C/O ratio of 1, we find that the molecular flux of CO is $\sim$13\% higher than in the model where oxygen was depleted to achieve a C/O ratio of 1 (see Fig.~\ref{CO_enrich} and Fig.~\ref{C/O_ratio_enrichC}). Enhancing carbon further to yield a C/O ratio of 3, there is even more emission from CO due to most of the available oxygen getting locked in CO; this also reduces the  \ce{H2O} abundance, resulting in less water line cooling, and thus higher gas temperatures and higher molecular fluxes (see black triangles in Fig.~\ref{C/O_ratio_enrichC}).

In general, the \ce{C2H2} flux is under-predicted relative to the observations when the C/O ratio is 1 and slightly over predicted when the C/O ratio is 3. In this latter case, we produce a pseudo-continuum of \ce{C2H2} with a stronger Q-branch relative to the pseudo-continuum, compared to the case of oxygen depletion.
Figure\,\ref{C2H2_enrich} (and Fig.\,\ref{C/O_ratio_enrichC}) shows the difference in the spectrum when C/O of 1 is attained either by depleting oxygen or by enriching carbon and a simultaneous
decrease in CO emission. Depletion of oxygen and enrichment of carbon thus lead to very different mid-IR spectra for the
same C/O ratio.
The rise in flux levels of both the molecules \ce{C2H2} and CO with increasing C abundance could be counteracted by increasing the dust-to-gas ratio. However, enhancing the C/O ratio with enriching the gas in carbon always increases the emission from \ce{C2H2} and CO simultaneously, because all the available oxygen is locked in CO. On the other hand, the depletion of oxygen allows an increase in molecular emission from \ce{C2H2} and a simultaneous decrease in CO emission. Thus, increasing carbon to enhance C/O does not suppress the CO molecular emission compared to \ce{C2H2}, while this is exactly what our disk models require to match the JWST observations.

\subsection{\ce{C2H2} emission}
\label{sect:C2H2}


\cite{Tabone2023} used two distinct components to explain the optically thin and thick emission of \ce{C2H2}. Our 2D disk model can explain the observations by a vertically and radially extended emitting region of \ce{C2H2}. When analysing the spectral range of only the Q-branch (13.5-13.76\,$\mu$m) in the model, we find similar column densities and emitting areas as reported in  Table\,\ref{TNR} for the entire \ce{C2H2} band, showing that the Q-branch dominates the statistical averages. 

\begin{figure}[h]
   \centering
    \includegraphics[width=\linewidth]{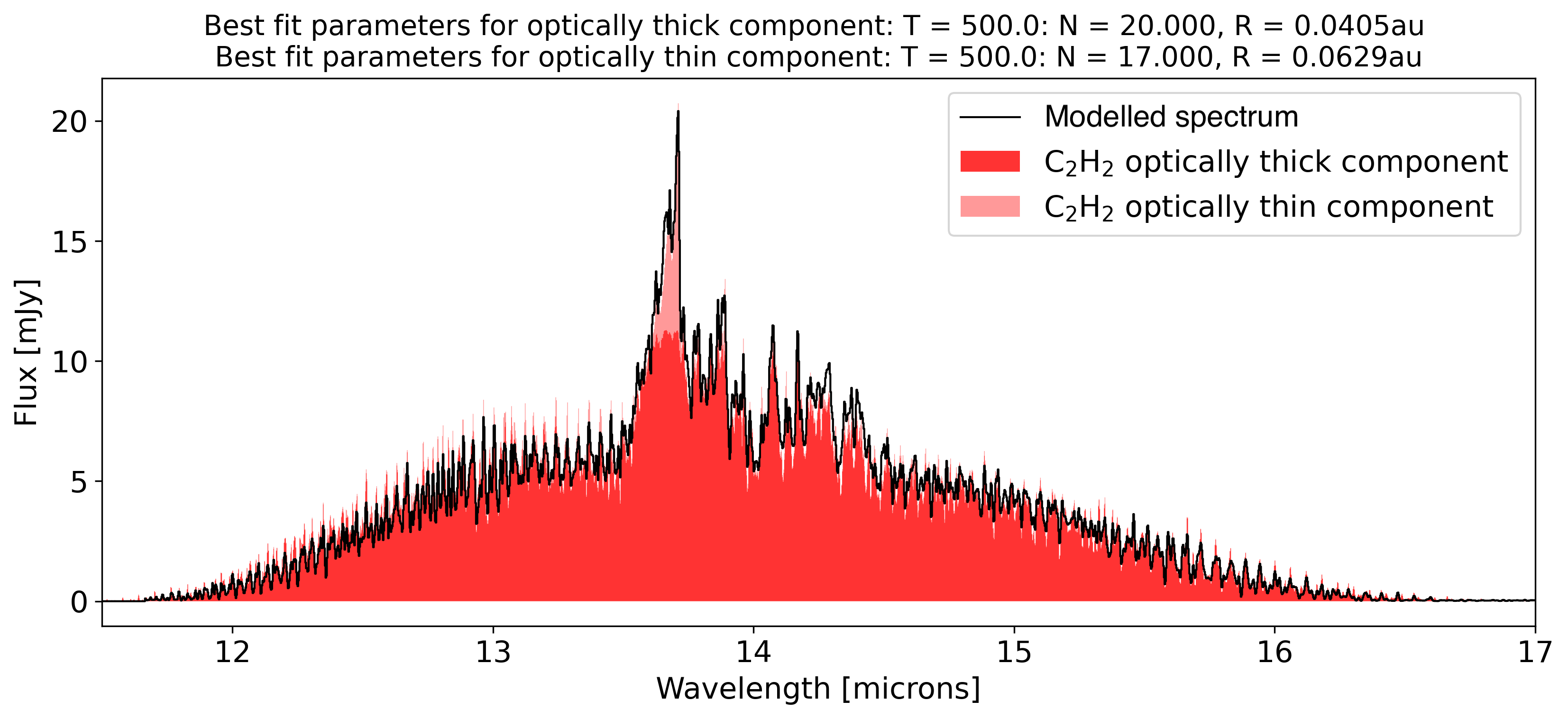}
   \caption{The 0D slab model fit to the 2D modeled spectrum of \ce{C2H2} which is convolved to a resolution of 2500 and resampled to JWST-MIRI/MRS. The light red and dark red colors depict the optically thin and thick component, respectively.}
              \label{slab_C2H2}%
\end{figure}

To investigate further how to bring the slab models and 2D disk models together, we fitted slab models to our simulated \ce{C2H2} spectrum (best model convolved to a resolution of 2500 and resampled). We use the slab models described in \cite{Arabhavi2023} and \cite{Kanwar2024}. We find that a single slab component is not able to produce the optically thin features of the simulated \ce{C2H2} band from the 2D model along with the pseudo-continuum. Hence, we also have to use two slab components. We fixed the temperature to $T\,=\,500$\,K for the optically thick and thin components and retrieve column densities of $N\,=\,10^{20}$ and $10^{17}$\,cm$^{-2}$ with emitting areas of 0.04 and 0.06\,au, respectively (Fig.~\ref{slab_C2H2}). These column densities are similar to those derived by \cite{Tabone2023} from JWST/MIRI-MRS observations using the same approach. We also find the optically thin component to be more radially extended than the thick component, similar to  \cite{Tabone2023}. So, the 2D emitting conditions of a thermo-chemical disk model can indeed be reproduced by a two component 0D slab model.

\subsection{\ce{C2H2} tracing UV luminosity}

We found earlier that our molecular fluxes react sensitively to the gas temperature. J160532 has an estimated mass accretion rate of $10^{-10}-10^{-9}$\,M$_\odot$\,yr$^{-1}$and we know that the mass accretion rate can be highly variable. This can result in varying levels of UV luminosity irradiating the disk.

\begin{figure}[h!]
   \centering
    \includegraphics[width=\linewidth]{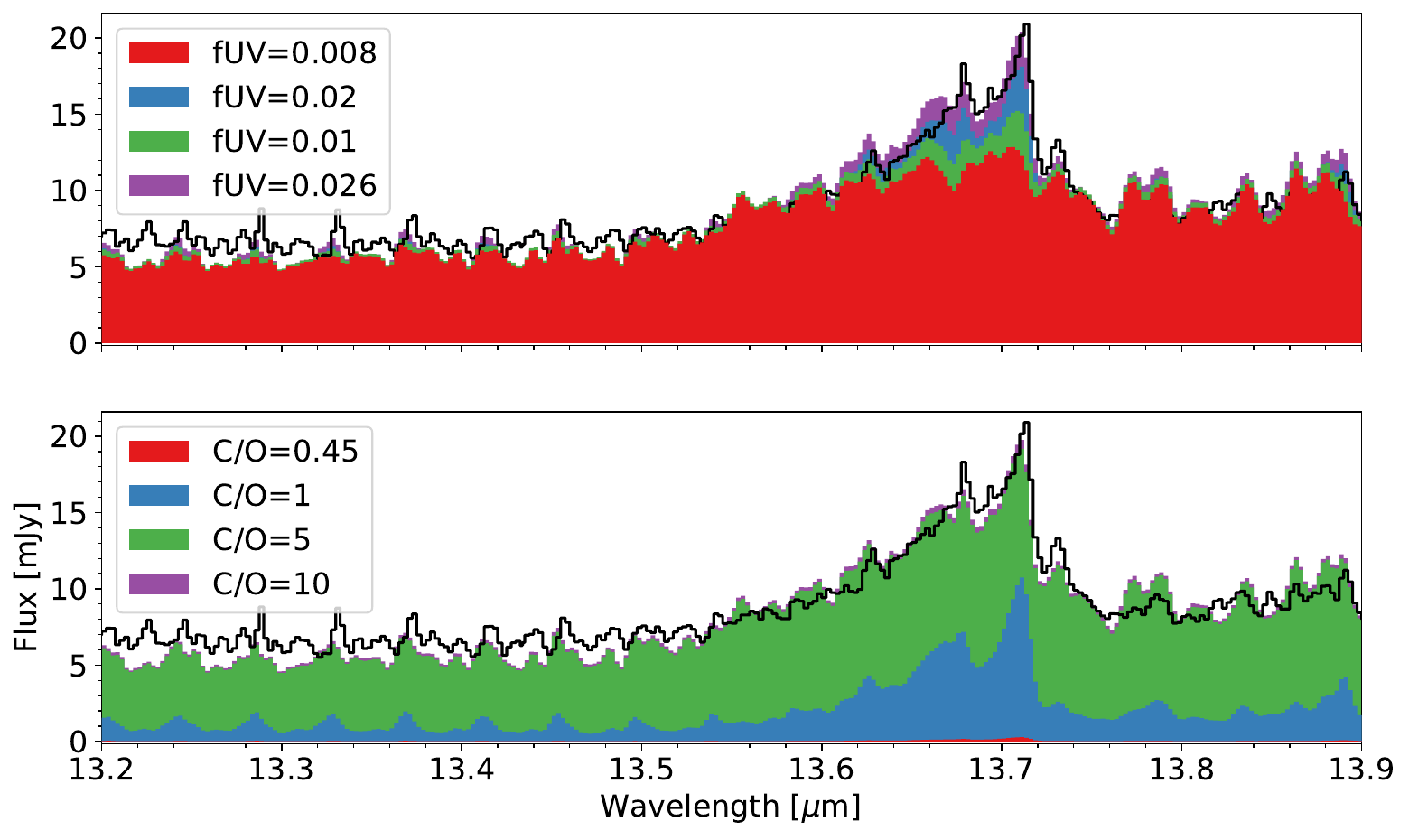}
   \caption{The effect of UV luminosity $L_{\rm{UV}}$ (upper panel) and C/O ratio (lower panel) on the \ce{C2H2} emission band at 13.5\,$\mu$m. The parameter fUV denotes the ratio between $L_{\rm{UV}}$ and $L_{\star}$. All models in the lower panel correspond to an $f_{\rm UV}$ value of 0.026.}
   \label{fUV}%
\end{figure}

\cite{Pascucci2013} inferred a UV luminosity of $2.5\,\times\,10^{-4}$\,$L_{\odot}$ for J160532 using a mass accretion rate of $\log M_{\rm acc}\,=\,-9.1$ and empirical scaling relations.  Variability and uncertainties in the extinction lead to a large scatter in such empirical relations and thus luminosity estimates \citep{Yang2012}. The model \ce{C2H2} emission using this UV luminosity ($f_{\rm{UV}}=\,L_{\rm UV}/L_\ast\,=\,0.008$) does not produce a visible Q-branch (red spectrum in Fig.\,\ref{fUV}, upper panel). However, when increasing $L_{\rm{UV}}$ by a factor $\sim\,3$ to $7.8\,\times\,10^{-4}$\,$L_{\odot}$ ($f_{\rm UV}\,=\,0.026$), we are able to reproduce the observed Q-branch. The strength of the Q-branch increases with $L_{\rm{UV}}$ similar to the findings of \cite{Woitke2024} for EX\,Lup. The Q-branch is generated by the optically thin surface layers of \ce{C2H2} whereas the weak lines in the P- and R-branch are emitted from deeper layers in the disk (see Fig.\,\ref{line_analysis}). Analysing the emitting conditions of the Q-branch separately, we find that the average gas temperature weighted with the abundance of \ce{C2H2} is lower in models with low $L_{\rm{UV}}$ relative to models with high $L_{\rm{UV}}$. Increasing the UV luminosity of the central star leads to an increase in the gas temperature, but only in the surface layers. 
As the P- and R-branch probe deeper layers, the flux from these lines changes very little. 

Increasing $L_{\rm{UV}}$ to $7.8\,\times\,10^{-4}$\,$L_{\odot}$ also impacts the flux and properties of molecules such as \ce{CO}, HCN and \ce{CH4}. The integrated flux of CO
increases by 39\% if the $f_{\rm UV}$ is increased from 0.008 to 0.026. In general, the molecular emission of these molecules decreases with increasing $L_{\rm{UV}}$. Our final best model uses this high UV luminosity and the emission of these two molecules is then consistent with the observations.

\section{Discussion}\label{dis}
We discuss in the subsequent paragraphs the scenarios for a very high C/O ratio, the caveats in the modelling and future observations.

\subsection{Scenarios for a very high C/O ratio}

Several earlier works investigated the impact of the C/O ratio on the disk chemistry and molecular emission \citep[e.g.,][]{Lee2010, Najita2011, Woitke2018, Wei2019, Anderson2021}. However, none of them considered values as extreme as 100. In the following, we briefly discuss the scenarios proposed in the literature and comment on whether they could lead to extreme C/O ratios.

We assume a disk gap both in the dust and the gas as mentioned in Sect.\,\ref{Modelling strategy}. The gap is crucial to produce the molecular emission from different species simultaneously. Gaps in disks can be formed by planet-disk interaction, but also by other mechanisms such as zonal flows \citep{Johansen2009} or dead zones \citep{Flock2015}. \cite{Gillon2017} reported extremely small semi-major axes for the seven planets found around TRAPPIST-1 (all orbit inside 0.07\,au). So our proposed gap could have been created by one or more planets. Which planet masses are required to open such a gap requires further investigation.

Our best disk model suggests that the inner disk is devoid of oxygen and not enhanced in carbon. One way to achieve this is when the water frozen on cold grains remains locked in the outer disk due to the presence of a deep gap. The second possibility is that the water-rich grains have already drifted inwards,  subsequently enriching the gas with oxygen and this gas has long accreted onto the central star. 
The timescales of dust growth and drift are very short in disks around very low-mass stars \citep{pinilla2023}. So, each of these scenarios could be equally likely depending on when and where the gap is formed. \cite{Kalyaan2023}, \cite{Mah2024} and \cite{Sellek2025a} studied the effect of the presence of a gap on pebble transport and the elemental abundances in the disk; \cite{Lienert2025} studied the influence of photoevaporation on the elemental abundances of the inner disk. \citet[][Fig.\,5]{Sellek2025b} reach a C/O$\,>\,$10 through depleting oxygen when an initial warm gap is present at 5\,au and a high cosmic ray ionization rate of $\zeta\,=\,10^{-17}$\,s$^{-1}$ which helps to convert CO into the less volatile \ce{CH4}. A C/O $\sim$5 is required within our model framework to produce a strong pseudo-continuum for \ce{C2H2}. However, a larger C/O ratio still produces formally an even better match for the entire MIRI spectrum. \cite{Mah2024} introduced a gap at 3\,au at the birth of the disk ($t\,=\,0$) around a solar-type star for a high viscosity disk ($\alpha\,=\,10^{-3}$) and find that the C/O becomes super-stellar after a span of 4\,Myr. However, the maximum C/O ratio they attained was only $\sim\,3$. The timescale to achieve such values could be much shorter for disks around VLMS due to the faster evolution. Our modeling also finds that the gap is located much closer to the star, between 0.1 and 0.5\,au. In addition, \cite{Mah2024} find that the time of gap formation affects the temporal evolution of C/O in the inner disk. So, we would need a more detailed modeling of the transport processes specifically in disks around VLMS including a wide range of gap locations, width and time of formation to fully explore the possible range of oxygen depletion.

Conversely, enrichment of carbon by irreversible destruction of refractory organics to \ce{C2H2} is explored in \cite{Houge2025}. They find C/O$\sim$1 for a disk viscosity of $\alpha = 10^{-3}$. If we enrich carbon in our thermo-chemical models so that the C/O ratio is 1, the line fluxes of \ce{C2H2} are lower than the observations as shown in Fig.\,\ref{C/O_ratio_enrichC}. We would require a ratio of 3 to get close to the observed \ce{C2H2} flux level.

Another way suggested leading to a high C/O ratio is the stellar accretion rate \citep{Colmenares2024}. The enrichment of carbon via sublimation leads to hydrocarbon chemistry. Sources with a high mass accretion rate such as Sz114 will accrete this C-rich gas onto the star, while sources with low accretion rates such as J160532 and Sz28 leave the disk enough time for the C-rich gas to be transported to the surface layers. It is still unclear what level of carbon enrichment can be reached in this way.

\subsection{Optically thin inner disk}

We find that to achieve very high column densities for \ce{C2H2}, a high C/O ratio along with a dust depleted inner disk is required. This implies that the inner disk is optically thin in dust. The fact that many disks lack silicate features can indicate that the grains are on average larger than a few micron; this leads to lower dust opacities compared to the commonly used dust grain size distribution. In addition, the detection of $^{13}$CCH$_2$ emission also indicates that deeper layers are being probed.

The specific disk structure we propose in this paper is able to explain the key spectral features observed in J160532. However, \cite{Arabhavi2025b} demonstrate also a wide chemical diversity among the disks around such very low-mass stars. An optically thin inner disk with a high C/O ratio could be a general feature to explain disks that show a pseudo-continuum of hydrocarbons in the MIRI spectra. However, not all aspects of the model can be generalized; for example, disks around Sz114 and 2MASS J15582981–2310077 may require alternative disk structures to explain the presence of strong \ce{H2O} emission. Thus, this work present a very first step in understanding the intricate disk structure information potentially encoded in these MIRI spectra.

\cite{Long2025} take the column density ratio of \ce{C2H2} and \ce{CO2} as a proxy for the C/O ratio and report a value of $\sim$1.5 for J160532. This approach assumes that these molecules share the same reservoir, and that the column densities are tracing similar regions in the disk. However, within our modeling framework, \ce{CO2} and \ce{C2H2} reside in different locations in the disk, the outer and inner disk respectively. 

\subsection{Caveats in the modelling}

We are able to explain the presence of the observed large column densities of hydrocarbons in the J160532 disk solely on the basis of the gas-phase chemistry and disk geometry. There is no ice formation in our inner disk due to high temperatures. The large grain sizes used in our model provide little opacity and hence UV radiation penetrates to the midplane preventing efficient ice formation even in colder environments. 
Thus, invoking surface chemistry will likely not cause changes in the determined parameters. 

Changing the absolute initial elemental abundances of C and O itself rather than the C/O ratio assuming solar initial elemental abundances can also lead to similar chemistry. However, such changes also lead to differences in the spectrum. A systematic approach of varying elemental abundances and studying its effects on mid-IR spectra is explored in \cite{Arabhavi2025c}.

ALMA observations are crucial for determining grain sizes in this disk, especially to inform the minimum grain size assumed for the outer disk. Small grains could provide more shielding for molecules such as \ce{CO2}. However, we did not see much effect on the \ce{CO2} and \ce{H2O} fluxes when using a larger minimum grain size (5\,$\mu$m) and a gas-to-dust mass ratio of 1000 in the outer disk. 

The heating and cooling processes are crucial to determine the gas temperature and this is closely intertwined with the chemistry. We include many such processes in the model \citep{Woitke2009, Woitke2024} and cooling rates are more complete for simple abundant molecules such as \ce{C2H2}, \ce{CH4}, \ce{CO}, \ce{CO2}, \ce{H2O}. However, when using a high C/O ratio, cooling by other hydrocarbons may also become crucial to reliably determine the gas temperature. For J160532, the entire mid-IR spectrum is dominated by \ce{C2H2} emission and its pseudo-continuum. The heating/cooling processes associated to this molecule are included in the model and therefore, the changes in the thermal structure may not be very large. The specific C/O ratio itself and how it affects the water cooling could have a larger effect compared to this. In fact, the water abundance and cooling in the surface layers reacts sensitively to our model parameters and small changes could lead to stronger water lines. 

The width of the Q-branch of \ce{CO2} is narrow in our best model due to it originating in low temperature gas. The overall \ce{CO2} flux is affected by parameters such as the C/O ratio and the gas-to-dust mass ratio in the outer disk along with the gap location. We show that the C/O in the outer disk cannot be smaller than $\sim\,$0.1 (such as 0.05) as then CO begins to emit from the outer disk with no dependency on the inner edge of the outer disk (see Fig.\,\ref{EA_lowC/O}). 
The C/O and gas-to-dust mass ratio in the outer disk are degenerate with the position of the inner edge of the outer disk. Within our explored parameters, we could not make the \ce{CO2} gas hot.

We show that the gas elemental abundances are crucial in determining the chemistry in the disk. This model assumes a set of elemental abundance (see Sect.\,\ref{Thermo-chemical modelling}), which are not self-consistently informed by any radial drift models (\citealt{Krijt2018}, \citeauthor{Mah2023}\, \citeyear{Mah2023}, \citeyear{Mah2024}), but rather change ad hoc at the gap location. More realistic 2D elemental abundance gradients informed by transport models could affect the detailed disk geometry required to reproduce the JWST spectrum as well as the gas-to-dust mass ratios.

\subsection{Future observations}

Overall, we lack strong observational constraints for the disk around J160532, but also VLMS disks in general. Future observations can aid to better constrain the disk parameters required in 2D disk modelling. The gas-mass used in our models is consistent with the estimates derived from the ALMA non-detection \citep{Barenfeld2016} and \ce{H2} lines observed with JWST-MIRI \citep{Riccardo2024}. In the best disk model, the continuum emission at 1.22\,mm is 0.035\,mJy, below the ALMA upper limit. So, deeper ALMA observations can help to better constrain the mass of the disk, at least the outer disk. ALMA observations in multiple bands could also constrain the grain sizes in the outer disk. Given how faint J160532 is, resolving the proposed gap structure with ALMA or ground-based instruments will remain a challenge. Upcoming facilities such as METIS/ELT will be able to spectrally resolve molecular lines and inform us about their emitting regions.
 

\section{Conclusions}\label{Conclusion}

We model the disk around the VLMS J160532 using the 2D thermo-chemical disk model P{\tiny RO}D{\tiny I}M{\tiny O}. We arrive at a structured disk model with a gap between 0.1 and 0.5\,au that closely matches the observed JWST-MIRI/MRS spectrum. 

\begin{itemize}
\item The model forms a molecular pseudo-continuum of \ce{C2H2} with a column density of $\sim$10$^{20}$cm$^{-2}$. This required an almost transparent inner disk, i.e.\ a low dust opacity with large grain sizes ($\geq\,5\,\mu$m) and settling. The detailed shape of the emission of \ce{C2H2} can be explained by a radially and vertically extended emitting region in our 2D disk model.

\item We find that the Q-branch of \ce{C2H2} probes the surface layers of the disk whereas the P- and R-branches trace the deeper layers of the disk. Hence, the Q-branch of \ce{C2H2} is sensitive to the UV luminosity $L_{\rm{UV}}$ of the central star in our modeling framework. $L_{\rm{UV}}$ affects mostly the gas temperature in the surface layers where the Q-branch originates. 

\item Within our modeling framework, the emission from molecules such as \ce{C2H2}, \ce{CH4}, HCN and CO originate from the inner disk whereas \ce{CO2} and \ce{H2O} emission is originating behind the inner wall of the outer disk.

\item The location of the gap is informed by the CO and \ce{CO2} fluxes. The gap is placed in a way to avoid any contribution of the outer disk to the CO flux while keeping \ce{CO2} hot enough to emit.

\item The presence of \ce{C2H2} along with CO is an indicator for the C/O ratio. We find that a large C/O ratio of $\sim$100 suppresses the CO emission lines and closely match the observations. In our modeling framework, a C/O ratio of 5 produces a molecular pseudo-continuum of \ce{C2H2} similar in strength to the observations.

\item We find that the combination of molecular fluxes can break degeneracies in elemental abundances. The depletion of oxygen to increase the C/O ratio can reduce CO emission and boost \ce{C2H2}, simultaneously, whereas the enrichment of carbon to increase the C/O ratio boosts fluxes of both molecules together.

\item We report the detection of the CO $\nu$= 2$\rightarrow$1 transition in the disk around J160532.
\end{itemize}

This work shows the immense potential of detailed 2D thermo-chemical disk models to analyse and interpret JWST-MIRI observations. While 0D slab models can provide much better overall fits to the molecular emission, the extracted emission conditions lack a physical and chemical context. Here, thermo-chemical models such as P{\tiny RO}D{\tiny I}M{\tiny O} can access additional information in these spectra that simpler approaches cannot. With this work we also demonstrate for the first time the potential of indirect spatial information encoded in such JWST-MIRI spectra from point sources. 

\begin{acknowledgements}
This work is based on observations made with the NASA/ESA/CSA \textit{James Webb} Space Telescope. The data were obtained from the Mikulski Archive for Space Telescopes at the Space Telescope Science Institute, which is operated by the Association of Universities for Research in Astronomy, Inc., under NASA contract NAS 5-03127 for JWST. These observations are associated with program \#1282. The following National and International Funding Agencies funded and supported the MIRI development: NASA; ESA; Belgian Science Policy Office (BELSPO); Centre Nationale d’Etudes Spatiales (CNES); Danish National Space Centre; Deutsches Zentrum fur Luft- und Raumfahrt (DLR); Enterprise Ireland; Ministerio De Econom\'ia y Competividad; Netherlands Research School for Astronomy (NOVA); Netherlands Organisation for Scientific Research (NWO); Science and Technology Facilities Council; Swiss Space Office; Swedish National Space Agency; and UK Space Agency. J.K., T.K., I.K., P.W. acknowledge support from the European Union’s Horizon 2020 research and innovation programme under the Marie Sklodowska-Curie grant agreement No.
860470 for this work. T.K. acknowledges support from STFC Grant ST/Y002415/1. YL acknowledges financial supports by the National Natural Science Foundation of China grant number 11973090, and by the International Partnership Program of Chinese Academy of Sciences grant number 019GJHZ2023016FN. B.T. is a Laureate of the Paris Region fellowship program, which is supported by the Ile-de-France Region and has received funding under the Horizon 2020 innovation framework program and Marie Sklodowska-Curie grant agreement No. 945298. T.H. acknowledge support from the European Research Council under the Horizon 2020 Framework Program via the ERC Advanced Grant Origins 83 24 28. E.v.D. and M.V. acknowledges support from the ERC grant 101019751 MOLDISK and the Danish National Research Foundation through the Center of Excellence ``InterCat'' (DNRF150). I.K., A.M.A., and E.v.D. acknowledge support from grant TOP-1 614.001.751 from the Dutch Research Council (NWO). D.B. has been funded by Spanish MCIN/AEI/10.13039/501100011033 grants PID2019-107061GB-C61 and No. MDM-2017-0737. We thank our referee Jenny K. Calahan for her constructive comments that help improve the paper.

\end{acknowledgements}
\bibliographystyle{aa}
\bibliography{Papers}

\appendix
\twocolumn
\section{Best model disk structure}\label{Properties of disk}

\begin{figure*}
   \centering
    \includegraphics[width=\linewidth]{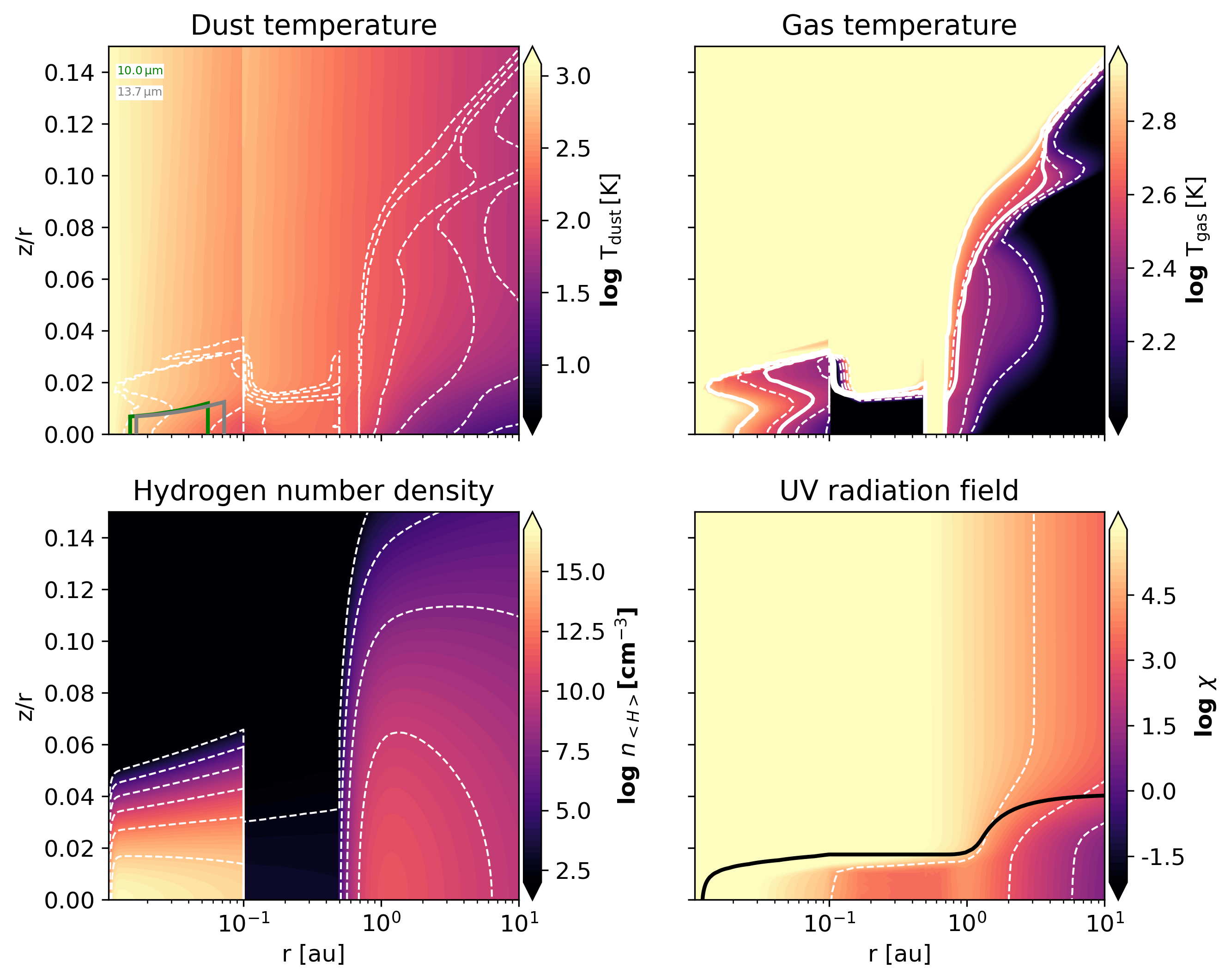}
   \caption{The 2D gas temperature, dust temperature, gas density and the UV radiation field structure in the best fit model. The white solid contours correspond to the gas temperature of 330 and 650\,K. The gray and green contours represent where the dust emission at 13.7 and 10\,$\mu$m is coming from. The black contour corresponds to the radial A$_{\rm{v}}$ of 1\,mag. The white dashed contours correspond to the value of the ticks on color bar. }
              \label{properties}%
\end{figure*}

Figure~\ref{properties} shows the thermal structure, gas density and UV radiation field in the best model. The dust emission at 10 and 13.7\,$\mu$m originating from deeper in the disk are also shown in green and gray. The disk is vertically optically thin due to the presence of large grains. The radial A$_{\rm{v}}$ of 1\,mag is shown in black. 

\begin{figure}
   \centering
    \includegraphics[width=\linewidth]{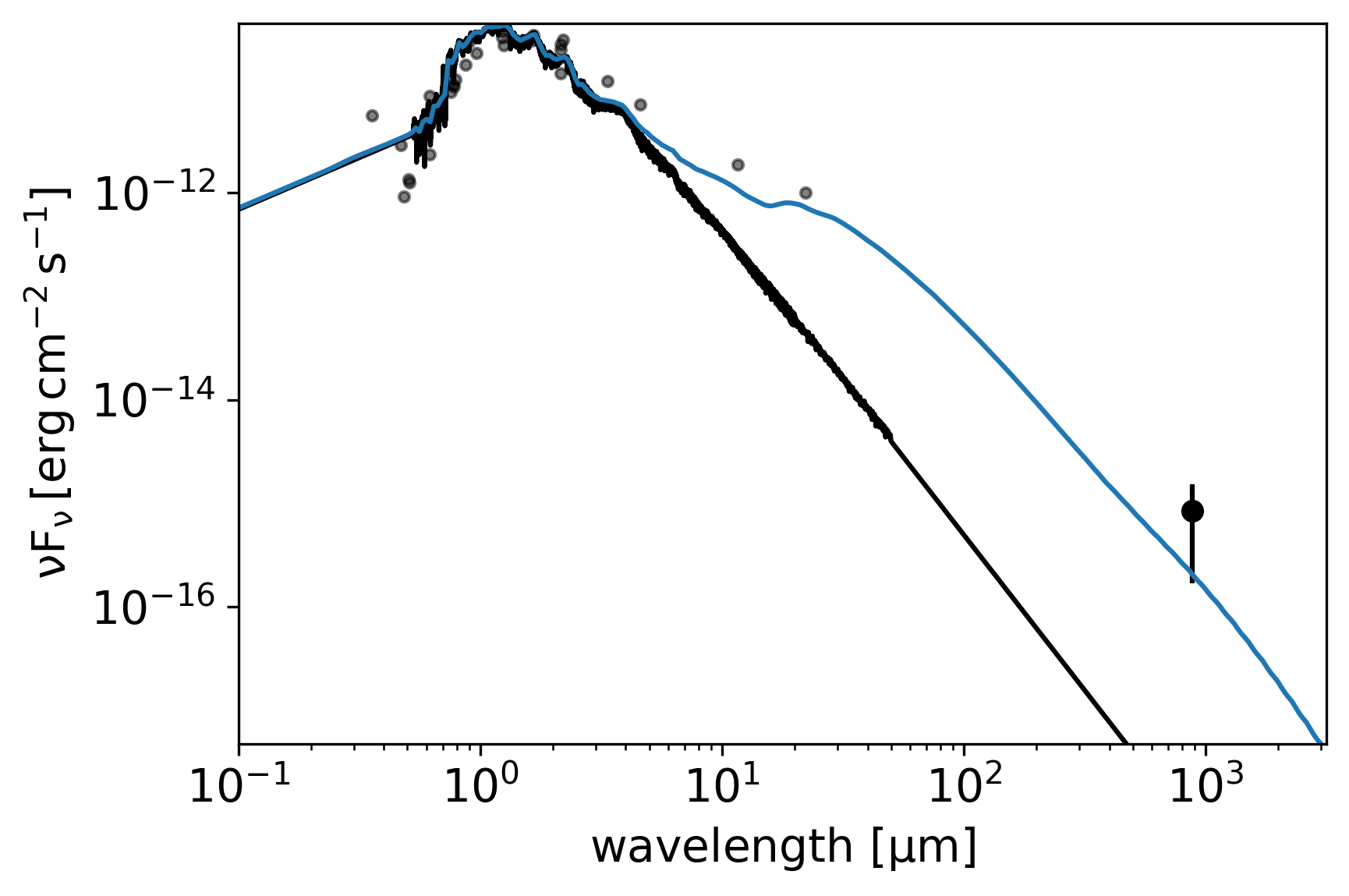}
   \caption{The SED of the final model in together with the photometric observations (black dots). The black line denotes the stellar contribution to the SED.}
              \label{SED}%
\end{figure}

The SED from the best-fit model along with the photometeric observations are shown in the Fig.~\ref{SED}. There is only an upper limit on the flux at long wavelengths obtained from \cite{Barenfeld2016} as this source was not detected with ALMA.

\section{Abundance of molecules}\label{abun}

\begin{figure*}
   \centering
    \includegraphics[width=\linewidth]{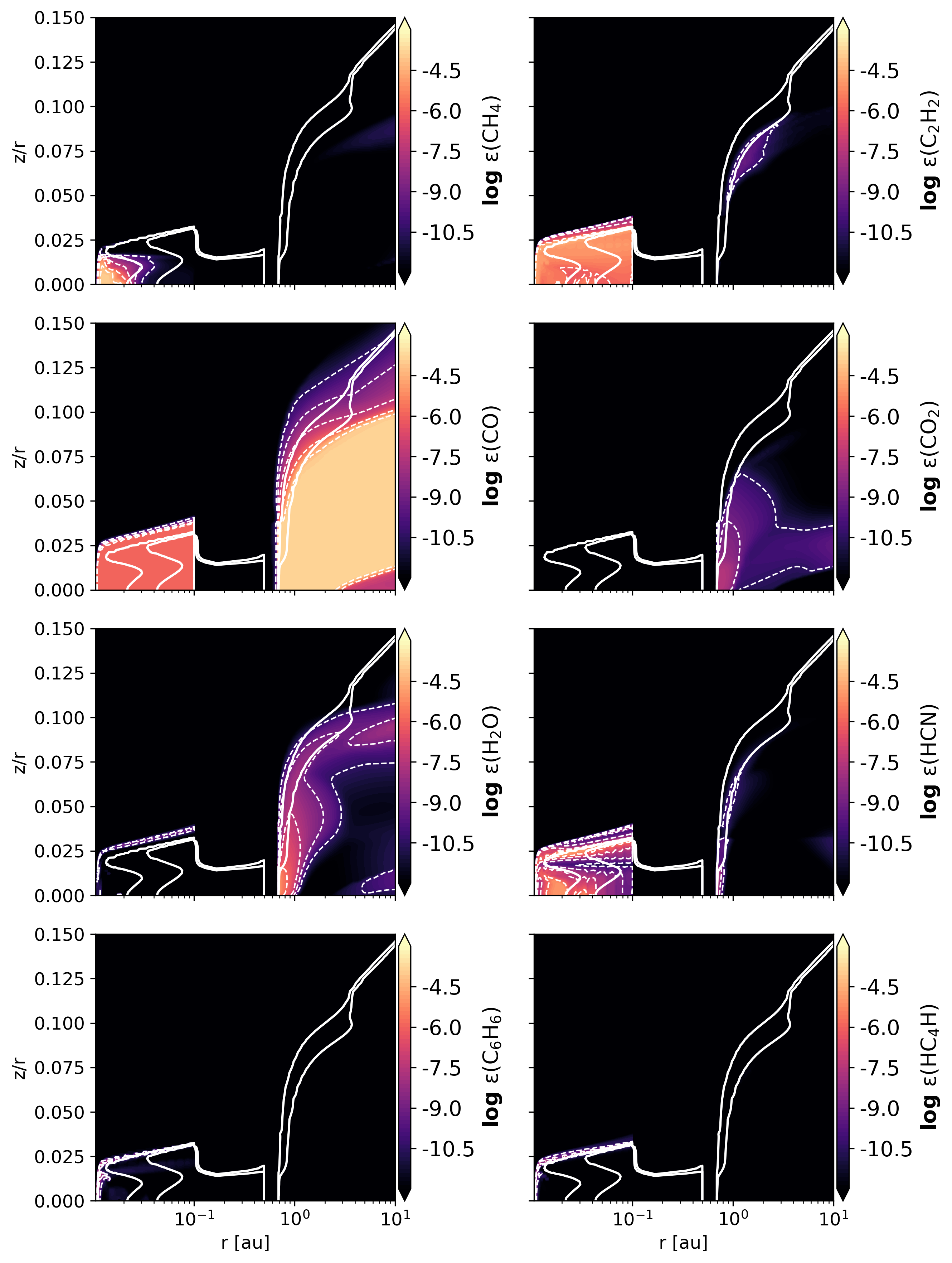}
   \caption{Abundances of different molecules in the best model. The solid white contours correspond to maximum and minimum temperatures 330 and 650\,K retrieved from slab model fitting in \cite{Tabone2023}. The dotted contours correspond to the value of the ticks on color bar.}
              \label{abundances}%
\end{figure*}

Figure~\ref{abundances} show the abundances of various molecules that are either detected or are used as a tool to determine the disk properties in this analysis. The inner disk has a high C/O ratio and thus is the reservoir of the hydrocarbons. The outer disk serves as an oxygen rich environment and molecules such as \ce{H2O} and \ce{CO2} reside in the outer disk.

\section{\ce{C2H2} line analysis}\label{Line analysis}

\begin{figure*}
   \centering
\includegraphics[width=\linewidth,height=200pt]{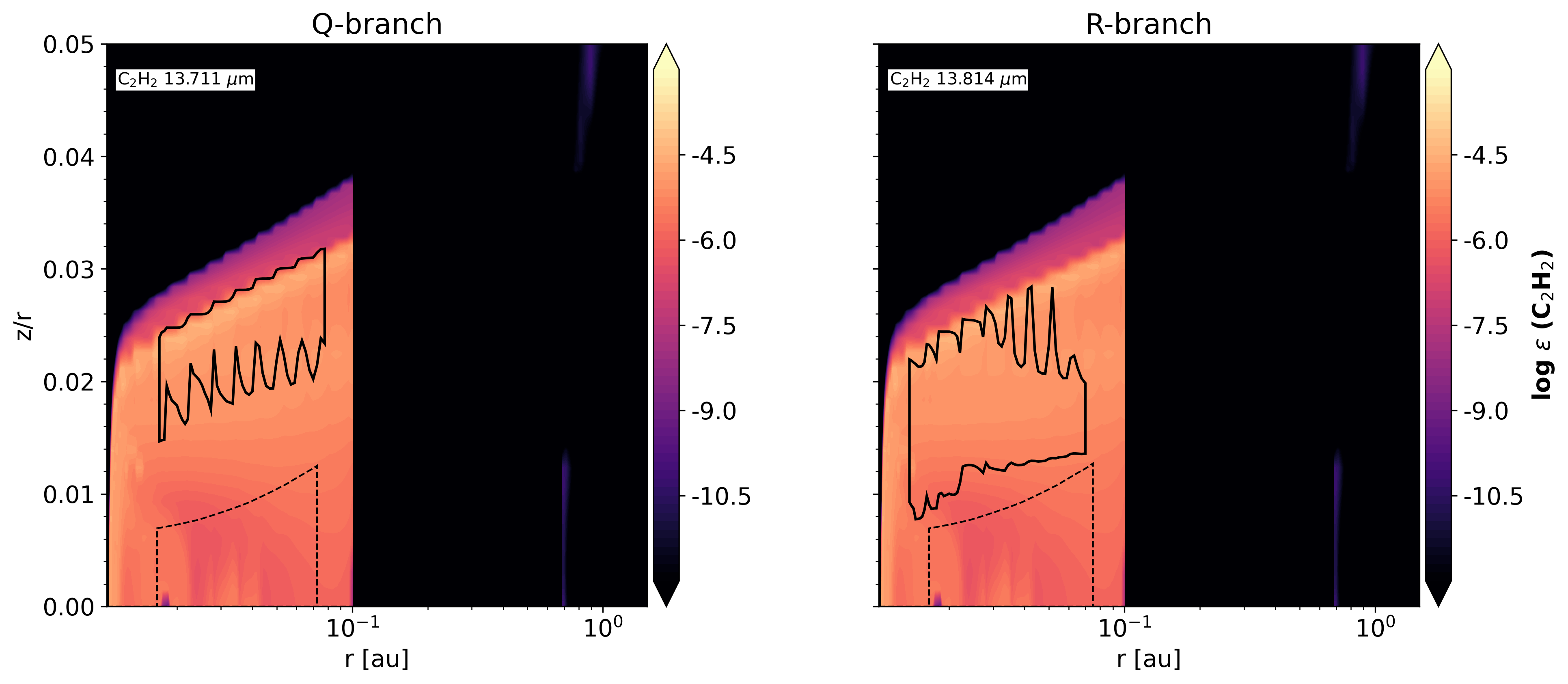}
   \caption{The abundance of \ce{C2H2} in the best model with the emitting area (solid black line) of a single strong molecular line from the Q- and R-branch spectral region of \ce{C2H2}. The black dashed contours denote the dust emission at that wavelength.}
              \label{line_analysis}%
\end{figure*}

Figure~\ref{line_analysis} shows the emitting regions of two different lines in the Q- and R-branch spectral region. We find that the average temperature weighted with the abundance of \ce{C2H2} in the emitting region of the R-branch is higher (493.22\,K) than that of the Q-branch (455.58\,K). This is because of the extended emitting region of the R-branch. 

On changing the $L_{\rm{UV}}$, the temperatures of the regions probed by the P- and R-branch change very little. The average gas temperature weighted by the abundance of \ce{C2H2} for the R-branch in the model with an $f_{\rm UV}$ of 0.008 decreases by $\sim$7\% relative to the final model, whereas for the Q-branch it decreases by $\sim$18\%. This shows that the effect of the $L_{\rm{UV}}$ on the gas temperature is more severe in the surface layers.

\section{Converging to the best model}\label{Different disk models}

\begin{figure}
   \centering
    \includegraphics[width=\linewidth]{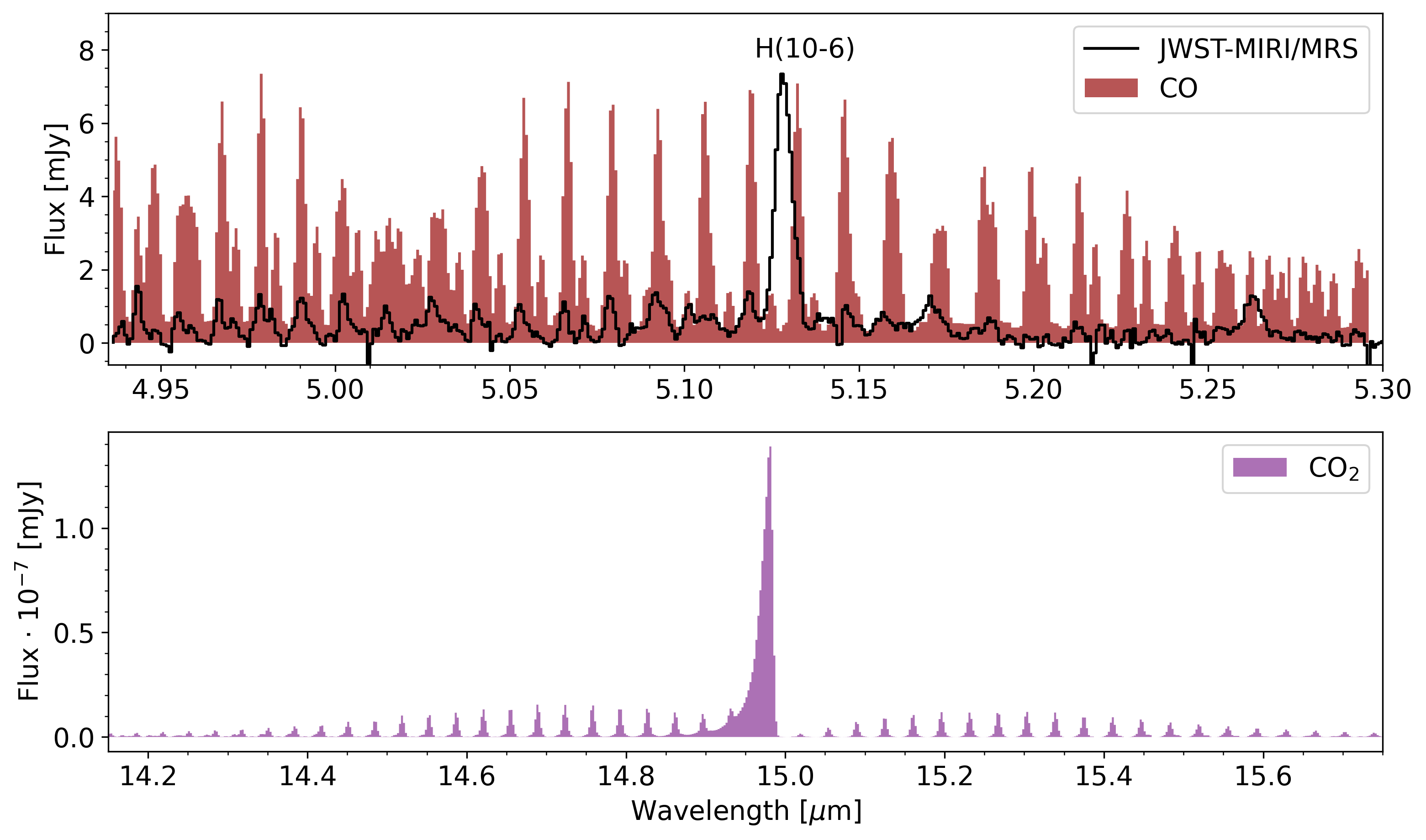}
   \caption{Molecular emission of \ce{CO} and \ce{CO2} from the model with a continuous disk structure convolved and resampled to JWST-MIRI/MRS resolution. The JWST-MIRI/MRS spectrum is shown in black.}
              \label{CO2_fulldisk}%
\end{figure}

We explored a range of disk parameters to identify a model that best reproduces the observed spectral features. In particular, we tested multiple approaches to simultaneously reproduce the \ce{C2H2} and \ce{CO2} emission features. Here, we present a selection of intermediate models considered during the process of converging on the final adopted model. A continuous disk structure is modelled using C/O\,=\,100 and a gas-to-dust mass ratio of 1000 along with dust settling. The modelled spectrum of \ce{CO} and \ce{CO2} produced by FLiTs, convolved to a spectral resolution of 3500 and 2500, respectively, and resampled to JWST-MIRI/MRS wavelength grid resolution is shown in Fig.~\ref{CO2_fulldisk}. The flux levels of \ce{CO2} are too low to be detected by JWST-MIRI. The CO flux is higher than our best-fit model even when we apply a huge oxygen depletion (C/O\,=\,100). This indicates the need of a gap to produce more \ce{CO2} flux and reduce the CO flux.

\begin{figure}
   \centering
    \includegraphics[width=\linewidth]{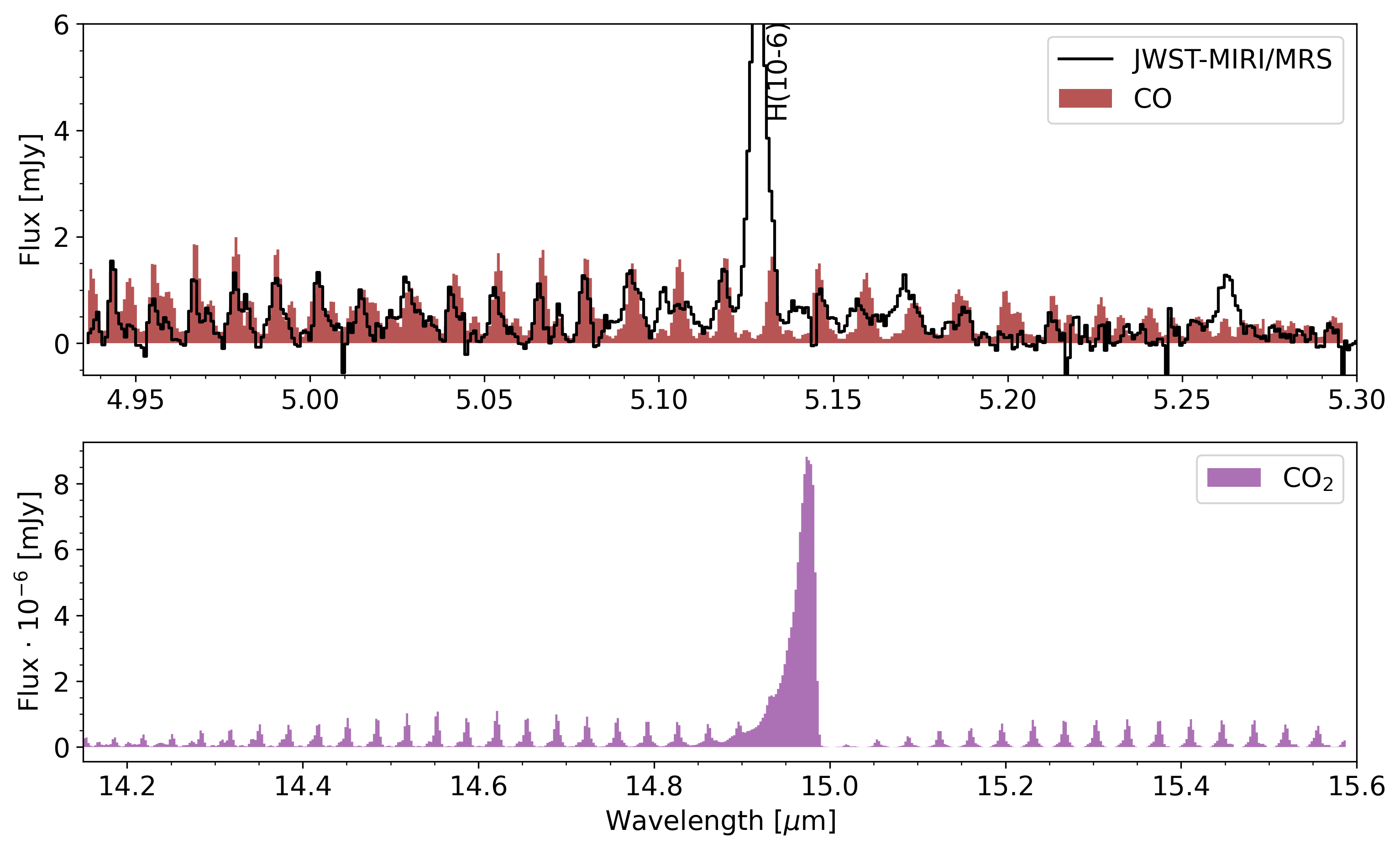}
   \caption{Molecular emission of \ce{CO} and \ce{CO2} from the model with a gap in the disk structure (C/O=100 in the inner and 10 in the outer disk) convolved and resampled to JWST-MIRI/MRS resolution. The JWST-MIRI/MRS spectrum is shown in black.}
              \label{CO2_gapdisk}%
\end{figure}

Figure~\ref{CO2_gapdisk} shows the molecular emission from CO and \ce{CO2} when there is a gap present in the disk. The inner disk has the parameters of the best model. The outer disk has a C/O ratio of 10. The modelled \ce{CO2} fluxes strongly underpredict the observations. This shows that an oxygen reservoir is required to generate detectable \ce{CO2} emission.

\begin{figure}
   \centering
    \includegraphics[width=\linewidth]{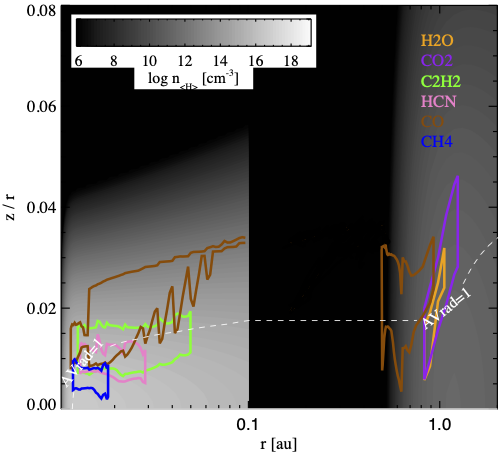}
   \caption{Emitting area of different molecules in the model when the C/O ratio is 0.05 in the outer disk. Contours are similar to Fig.~\ref{EA}.}
              \label{EA_lowC/O}%
\end{figure}

Figure~\ref{EA_lowC/O} illustrates the emitting area of different species in the model where all the parameters match the best model except for the C/O ratio in the outer disk. When the C/O ratio in the outer disk is higher than 0.1 (extremely O-rich, such as 0.05), CO emits from the outer disk irrespective of the location of the gap.

\section{Trends}

\begin{figure}
   \centering
    \includegraphics[width=\linewidth]{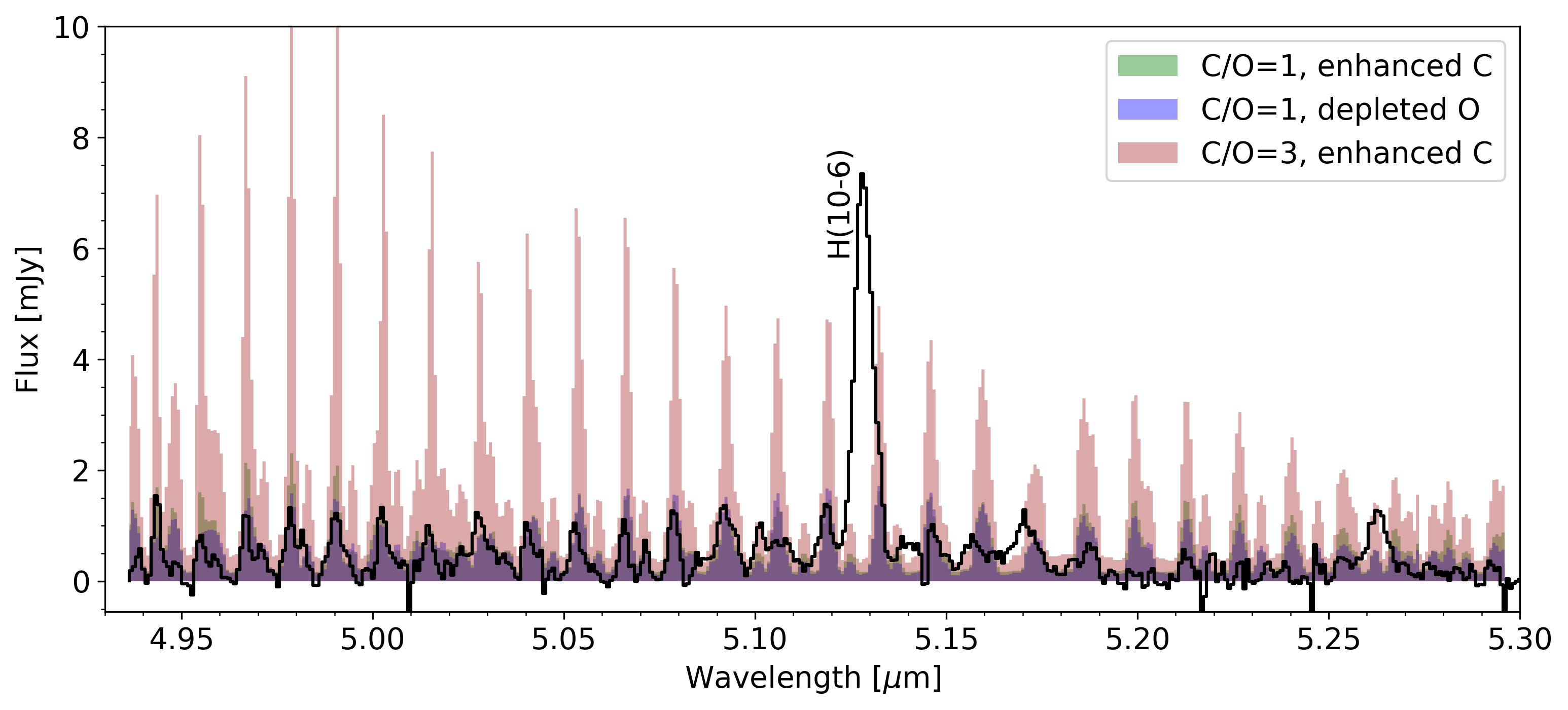}
   \caption{Molecular flux of CO in models with a C/O ratio of 1 and 3 attained by enriching carbon and a C/O ratio of 1 attained by depleting oxygen.}
              \label{CO_enrich}%
\end{figure}

\begin{figure}
   \centering
    \includegraphics[width=\linewidth]{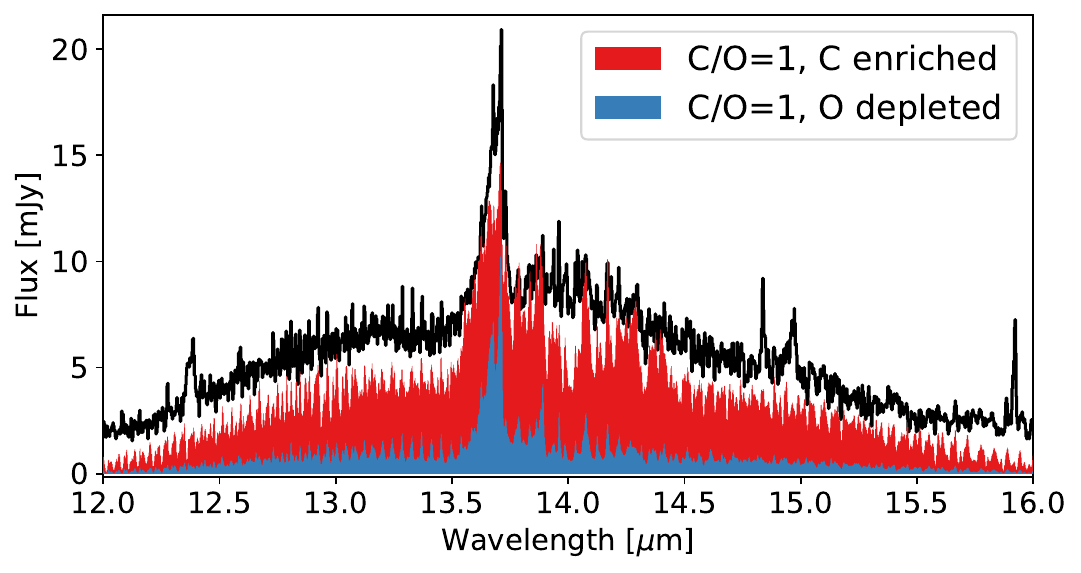}
   \caption{Molecular flux of \ce{C2H2} in models with a C/O ratio of 1 attained by enriching carbon and by depleting oxygen.}
              \label{C2H2_enrich}%
\end{figure}

\begin{figure}
   \centering
    \includegraphics[width=\linewidth]{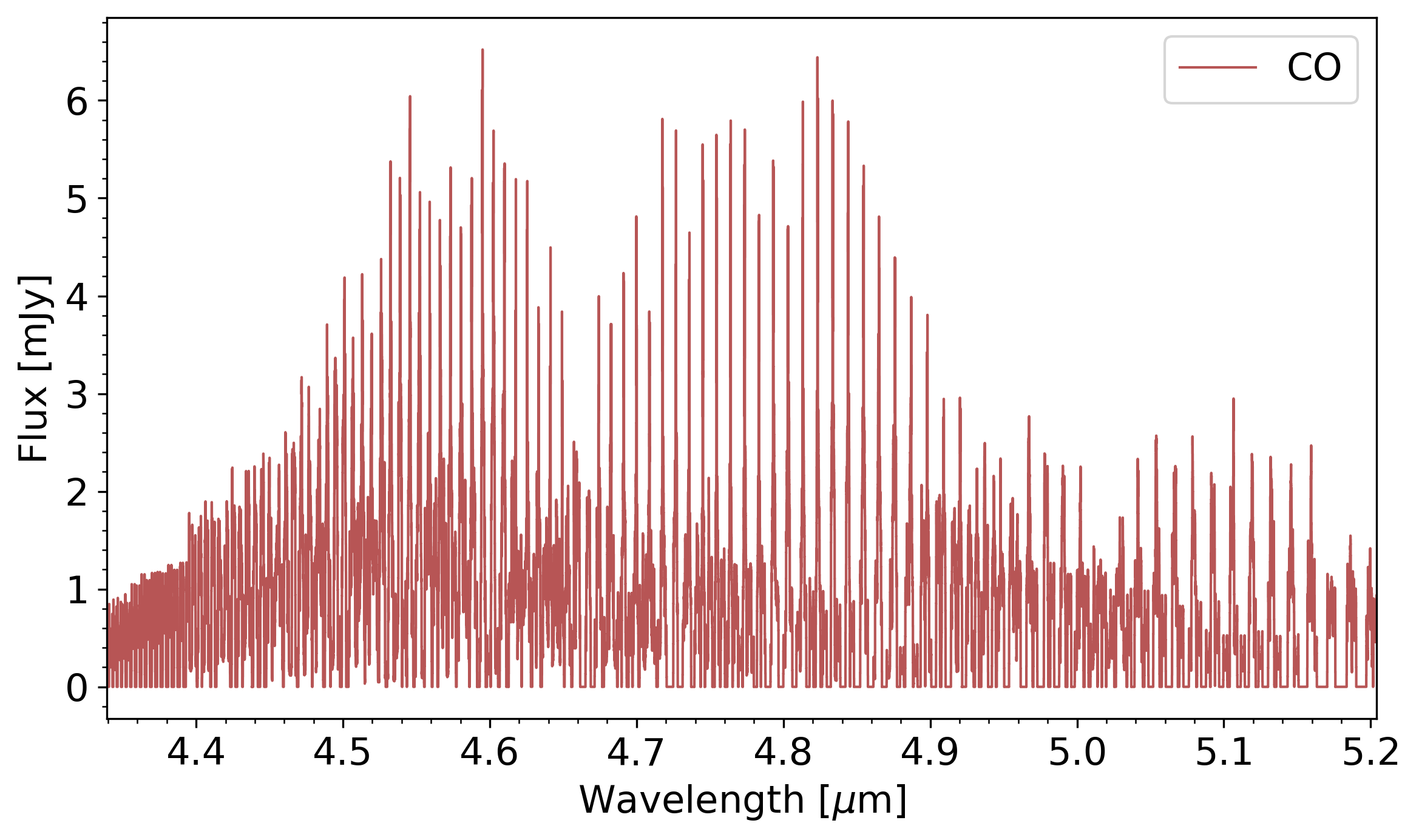}
   \caption{The emission of $^{12}$CO convolved to a resolution of 2700 from the best model.}
              \label{CO_nirspec}%
\end{figure}

Figure~\ref{CO_enrich} depicts the change in spectrum of CO when C/O ratio is enhanced either by depleting oxygen or enriching carbon in the gas in the disk. Figure\,\ref{C2H2_enrich} depicts the change in spectrum of \ce{C2H2} when  C/O ratio of 1 is attained by depleting oxygen and enriching carbon.
Figure~\ref{CO_nirspec} shows the modelled CO flux at NIRSpec wavelength region convolved to a resolution of 2700. We did not consider here the isotopologues such as $^{13}$CO, C$^{18}$O and C$^{17}$O.


\begin{figure}
   \centering
    \includegraphics[width=\linewidth]{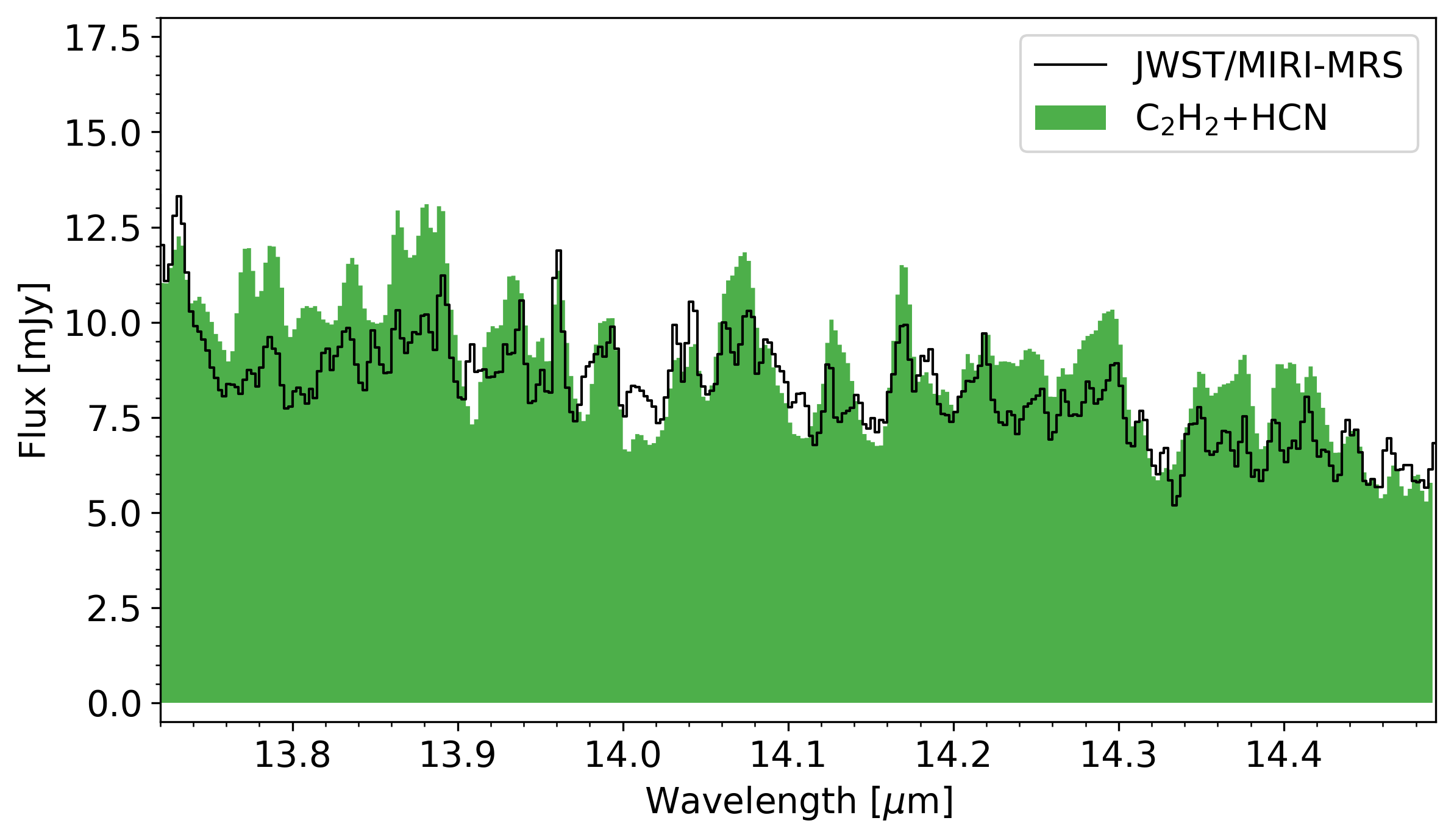}
   \caption{The spectrum of \ce{C2H2} and HCN together considering opacity overlap on top of the observed JWST-MIRI/MRS spectrum in black.}
              \label{C2H2+HCN}%
\end{figure}


%
%
\end{document}